\documentclass[epj,nopacs]{svjour}

\usepackage{amssymb}
\usepackage{epsfig}
\usepackage{eucal}
\usepackage{amsmath}

\newcommand{\ta}[1]{#1\hspace{-.42em}/\hspace{-.07em}} 
\newcommand{\beq}{\begin{equation}}
\newcommand{\eeq}{\end{equation}}
\newcommand{\bea}{\begin{eqnarray}}
\newcommand{\eea}{\end{eqnarray}}
\newcommand{\non}{\nonumber}
\newcommand{\eq}[1]{eq.(\ref{#1})}

\newcommand{\be}{\beta}

\headnote{
\hfill \parbox{4cm}{\tt \normalsize CERN-TH/2001-350 \\ TTP01-32 \\}}

\title{Radiative return at NLO and the measurement of the \\ 
hadronic cross-section in electron--positron annihilation}

\author{Germ\'an Rodrigo\inst{1}\thanks{Supported in part by 
E.U. TMR grant HPMF-CT-2000-00989; \email{german.rodrigo@cern.ch}} \and
Henryk Czy\.z\inst{2,3}\thanks{Supported in part by 
EC 5-th Framework, contract HPRN-CT-2000-00149; \email{czyz@us.edu.pl}}\and
Johann H. K\"uhn\inst{1,4}\thanks{\email{jk@particle.uni-karlsruhe.de}}
 \and Marcin Szopa\inst{2} }

\institute{TH-Division, CERN, CH-1211 Geneva 23, Switzerland. \and 
Institute of Physics, University of Silesia,
PL-40007 Katowice, Poland. \and
Institute of Advanced Study, University of Bologna, I-40138 Bologna,Italy \and
Institut f\"ur Theoretische Teilchenphysik,
Universit\"at Karlsruhe, D-76128 Karlsruhe, Germany.}

\date{Received: \today}

\abstract{Electron--positron annihilation into hadrons plus an 
energetic photon from initial state radiation allows the hadronic 
cross-section to be measured over a wide range of energies.
The full next-to-leading order QED corrections for the 
cross-section for $e^+ e^-$ annihilation into a real tagged photon 
and a virtual photon converting into hadrons are calculated where 
the tagged photon is radiated off the initial electron or positron. 
This includes virtual and soft photon corrections 
to the process $e^+ e^- \rightarrow \gamma+\gamma^*$
and the emission of two real hard photons:
$e^+ e^- \rightarrow \gamma+\gamma+\gamma^*$.
A Monte Carlo generator has been constructed, which incorporates 
these corrections and simulates the production of two charged 
pions or muons plus one or two photons.
Predictions are presented for centre-of-mass energies between
$1$ and $10$~GeV, corresponding to the energies of DA$\Phi$NE, CLEO-C 
and $B$-meson factories.}

\begin{document}

\authorrunning{G.~Rodrigo, H.~Czy\.z, J.H.~K\"uhn and M.~Szopa}
\titlerunning{Radiative return at NLO}
\maketitle

\section{Introduction}

Electroweak precision measurements have become one of the central
issues in particle physics nowadays. The recent measurement of the 
muon anomalous magnetic moment $a_{\mu} \equiv (g-2)_{\mu}/2$ at 
BNL~\cite{Brown:2001mg} led to a new world average, differing by
$2.6$ standard deviations from its theoretical Standard Model 
evaluation~\cite{Hughes:1999fp}.
This disagreement, which has been taken as an indication of new physics,
has triggered a raving and somehow controversial deluge of non-Standard 
Model explanations. A new measurement with an accuracy three times 
better is under way; this will challenge even more the 
theoretical predictions.

One of the main ingredients in the theoretical prediction 
for the muon anomalous magnetic moment
is the hadronic vacuum polarization contribution~\cite{hadronicmuon}
which is moreover responsible for the bulk of the theoretical error.
It is in turn related via dispersion relations to the 
cross-section for electron--positron annihilation into hadrons,
$\sigma_{had}=\sigma(e^+ e^- \rightarrow hadrons)$. 
This quantity plays an important role also in the 
evolution of the electromagnetic coupling $\alpha_{QED}$ from the 
low energy Thompson limit to high energies~\cite{hadronicmuon,runningQED}.
This indeed means that the interpretation of improved measurements 
at high energy colliders such as LEP, the LHC, or the Tevatron depends 
significantly on the precise knowledge of $\sigma_{had}$.

The evaluation of the hadronic vacuum polarization
contribution to the muon anomalous magnetic moment, and even more so
to the running of $\alpha_{QED}$, requires the measurement of
$\sigma_{had}$ over a wide range of energies. Of particular 
importance for the QED coupling at $M_Z$ is the low energy region 
around $2$~GeV, where $\sigma_{had}$ is at present poorly determined 
experimentally and only marginally consistent with the 
predictions based on pQCD. New efforts are therefore 
mandatory in this direction, which could help to either 
remove or sharpen the discrepancy between theoretical 
prediction and experimental results for $(g-2)_{\mu}$
and provide the basis for future more precise high energy 
experiments.

The feasibility of using tagged photon events at high luminosity
electron--positron storage rings, such as the $\phi$-factory DA$\Phi$NE
or $B$-factories, to measure $\sigma_{had}$ over a wide range of 
energies has been proposed and studied in detail 
in~\cite{Binner:1999bt,Melnikov:2000gs,Czyz:2000wh} 
(see also~\cite{Spagnolo:1999mt,Khoze:2001fs}).
In this case, the machine is operating at a fixed centre-of-mass (cms)
energy. Initial state radiation (ISR) is used to reduce the effective energy 
and thus the invariant mass of the hadronic system. The measurement 
of the tagged photon energy helps to constrain the kinematics,
which is of particular importance for multimeson final states. 
In contrast to the conventional 
energy scan~\cite{Akhmetshin:1999uj}, the systematics of the 
measurement (e.g. normalization, beam energy) have to be taken into 
account only once, and not for each individual energy point independently. 

Radiation of photons from the hadronic system (final state radiation, FSR) 
should be considered as background and can be suppressed by choosing 
suitable kinematical cuts, or controlled by the simulation, once a 
suitable model for this amplitude has been adopted. 
From studies with the generator EVA~\cite{Binner:1999bt} one finds that 
selecting events with the tagged photons close to the beam axis and well 
separated from the hadrons indeed reduces FSR to a reasonable limit. 
Furthermore, the suppression of FSR overcomes the problem of its model 
dependence, which should be taken into account in a completely inclusive 
measurement~\cite{Hoefer:2001mx}.

Preliminary experimental results using this method have 
been presented recently by the KLOE collaboration at 
DA$\Phi$NE~\cite{Aloisio:2001xq,Denig:2001ra,Adinolfi:2000fv}.
Large event rates were also observed by the BaBar 
collaboration~\cite{babar}.

In this paper we consider the full next-to-leading order (NLO) QED
corrections to ISR in the annihilation 
process $e^+ e^- \rightarrow \gamma + hadrons$ where the photon is 
observed under a non-vanishing angle relative to the beam direction. 
The virtual and soft photon corrections~\cite{Rodrigo:2001jr}
and the contribution of the emission of a second real hard photon 
are combined to obtain accurate predictions for the exclusive 
channel $e^+ e^- \rightarrow \pi^+ \pi^- \gamma$ at cms energies 
of $1$ to $10$~GeV, corresponding to the energies of DA$\Phi$NE,
CLEO-C and $B$-meson factories. An improved Monte Carlo generator,
denoted PHOKHARA, includes these terms and will be presented in this work. 
The comparison with the EVA~\cite{Binner:1999bt} Monte Carlo, which simulates 
the same process at leading order (LO) and includes additional collinear 
radiation through structure function (SF) techniques, is described.
Predictions are presented also for the muon pair production channel
$e^+ e^- \rightarrow \mu^+ \mu^- \gamma$, which is also simulated
with the new generator. 

\section{Virtual and soft photon corrections to ISR}

\label{sec:virtsoft}

\begin{figure}
\begin{center}
\epsfig{file=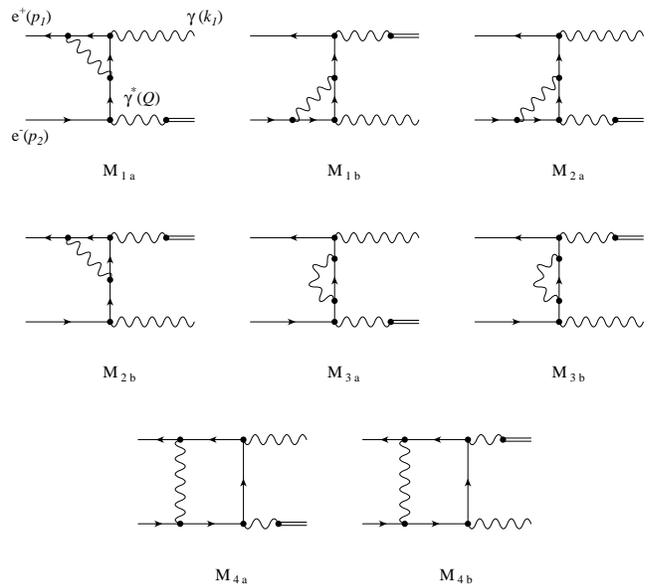,width=8.5cm}
\end{center}
\caption{One-loop corrections to initial state radiation in 
$e^+ e^- \rightarrow \gamma + hadrons$.}
\label{fig:nlo}
\end{figure}

At NLO, the $e^+ e^-$ annihilation process
\begin{align}
e^+(p_1) + e^-(p_2) \rightarrow & \gamma^*(Q) + \gamma(k_1)~,
\end{align}
where the virtual photon converts into a hadronic final state,
$\gamma^*(Q) \rightarrow hadrons$, and the real one is emitted
from the initial state, receives contributions from one-loop 
corrections (see Fig.~\ref{fig:nlo}) and from the 
emission of a second real photon (see Fig.~\ref{fig:real}). 

After renormalization the one-loop matrix elements still contain  
infrared divergences. These are cancelled by adding the contribution
where a second photon has been emitted from the initial state.
This rate is integrated analytically in phase space 
up to an energy cutoff $E_{\gamma}<w\sqrt{s}$ far below $\sqrt{s}$. 
The sum~\cite{Rodrigo:2001jr} is finite; however, it depends 
on this soft photon cutoff. The contribution from the emission of 
a second photon with energy $E_{\gamma}>w\sqrt{s}$ completes the 
calculation and cancels this dependence. 

In order to facilitate the extension of the Monte Carlo simulation 
to different hadronic exclusive channels the differential rate is 
cast into the product of a leptonic and a hadronic tensor and the 
corresponding factorized phase space:
\begin{equation}
d\sigma = \frac{1}{2s} L_{\mu \nu} H^{\mu \nu}
d \Phi_2(p_1,p_2;Q,k_1) d \Phi_n(Q;q_1,\cdot,q_n) 
\frac{dQ^2}{2\pi}~,
\end{equation}
where $d \Phi_n(Q;q_1,\cdot,q_n)$ denotes the hadronic 
$n$-body phase space including all statistical factors 
and $Q^2$ is the invariant mass of the hadronic system.

The physics of the hadronic system, whose description is 
model-dependent, enters only through the hadronic tensor 
\begin{equation}
H^{\mu \nu} = J^{\mu} J^{\nu +}~,
\end{equation}
where the hadronic current has to be parametrized through form factors.
For two charged pions in the final state, the current 
\begin{equation}
J^{\mu}_{2\pi} = i e F_{2\pi}(Q^2) \; (q_{\pi^+}-q_{\pi^-})^{\mu}
\label{hc}
\end{equation}
is determined by only one function, the pion form factor 
$F_{2\pi}$~\cite{Kuhn:1990ad}.
The hadronic current for four pions exhibits a more complicated 
structure and has been discussed in~\cite{Czyz:2000wh}.

The leptonic tensor, which describes the NLO 
virtual and soft QED corrections to initial state radiation in 
$e^+ e^-$ annihilation, has the following general form: 
\begin{align}
L^{\mu \nu}_{\mathrm{virt+soft}} &=
\frac{(4 \pi \alpha)^2}{Q^4 \; y_1 \; y_2} \;
\bigg[ a_{00} \; g^{\mu \nu} + a_{11} \; \frac{p_1^{\mu} p_1^{\nu}}{s}
 + a_{22} \; \frac{p_2^{\mu} p_2^{\nu}}{s} \non \\
&+ a_{12} \; \frac{p_1^{\mu} p_2^{\nu} + p_2^{\mu} p_1^{\nu}}{s}
+ i \pi \; a_{-1} \; 
\frac{p_1^{\mu} p_2^{\nu} - p_2^{\mu} p_1^{\nu}}{s} \bigg]~,
\label{generaltensor}
\end{align}
where $y_i=2k_1\cdot p_i/s$.
The scalar coefficients $a_{ij}$ and $a_{-1}$ allow the following expansion 
\begin{equation}
a_{ij} = a_{ij}^{(0)} + \frac{\alpha}{\pi} \; a_{ij}^{(1)}~, \qquad
a_{-1} = \frac{\alpha}{\pi} \; a_{-1}^{(1)}~,
\end{equation}
where $a_{ij}^{(0)}$ give the LO contribution. 
The NLO coefficients $a_{ij}^{(1)}$ and $a_{-1}^{(1)}$ were calculated
in~\cite{Rodrigo:2001jr} (see also~\cite{Khoze:2001fs})
for the case where the tagged photon is far from the collinear region.

The expected soft and collinear behaviour of the leptonic 
tensor is manifest when the following expression is used:
\begin{align}
L^{\mu\nu}_{\mathrm{virt+soft}} &= 
\frac{1}{1-\delta_{VP}} \bigg[ L_0^{\mu\nu} \bigg\{1 +
\frac{\alpha}{\pi}  \bigg[ -\log(4w^2) [1+\log(m^2)]  \non \\ 
&- \frac{3}{2} \log(m^2) - 2 + \frac{\pi^2}{3} \bigg] \bigg\}
+ C^{\mu\nu} \bigg]~,
\label{nloleptonic}
\end{align}
where $L_0^{\mu\nu}$ stands for the LO leptonic tensor. 
The first term, $\log(4w^2) [1+\log(m^2)]$ where $m^2=m_e^2/s$, 
contains the dependence on the soft photon cutoff $w$, which has to 
be cancelled against the contribution from hard radiation.
The next three terms, also proportional to the LO leptonic tensor, 
represent the QED corrections in leading log approximation with 
the typical logarithmic dependence on the electron mass.
The tensor $C^{\mu \nu}$ finally contains the subleading QED corrections. 

The factor $1/(1-\delta_{VP})$ accounts for the vacuum polarization 
corrections in the virtual photon line. This multiplicative correction 
can be approximately reabsorbed by a proper choice of the running 
coupling constant.
In the present version of the Monte Carlo generator one can choose
to include or not the contribution from the real part of the lowest 
order leptonic loops~\cite{BHS:1986}:
\begin{align}
\delta_{VP}(q^2) = \frac{\alpha}{3\pi} \sum_{i=e,\mu,\tau} 
 \biggl[\frac{1}{3}-\biggl(1+\frac{2m_i^2}{q^2}\biggr)F(q^2,m_i)\biggr]
 \ \ ,
\end{align}
where
\begin{equation}
F(q^2,m) = 
  \begin{cases}
  2 + \beta\log\left(\frac{1-\beta}{1+\beta}\right) \ , &
  \ \mbox{\rm for}  \ \ q^2 > 4m^2 \ , \\
  2 - \bar \beta \arctan\left(\frac{1}{\bar\beta}\right) \ , &
  \ \mbox{\rm for}  \ \ 0< q^2 < 4m^2 \ ,
   \end{cases}
\end{equation}
with
\begin{eqnarray}
\beta = \sqrt{1-\frac{4m^2}{q^2}} \  \ \mbox{\rm and} \ \ 
 \bar \beta = \sqrt{\frac{4m^2}{q^2}-1} \ \ .
\end{eqnarray}
This routine can be easily replaced by a user if necessary.

\section{Emission of two hard photons}

\label{sec:two}

\begin{figure}
\begin{center}
\epsfig{file=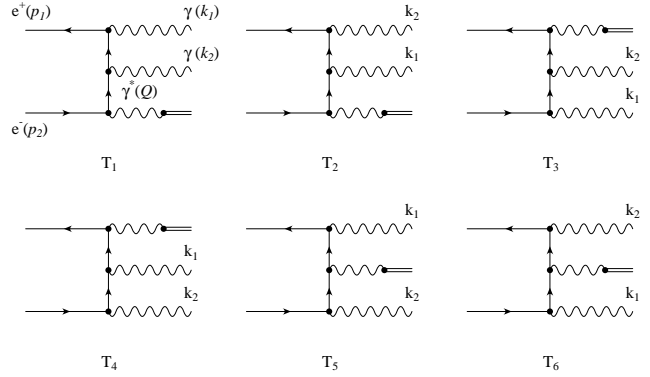,width=8.5cm}
\caption{Emission of two real photons from the initial state 
in $e^+e^-$ annihilation into hadrons.}
\label{fig:real}
\end{center}
\end{figure}

In this section, the calculation of the matrix elements for the 
emission of two real photons from the initial state 
\begin{align}
e^+(p_1) + e^-(p_2) \rightarrow  \gamma^*(Q) + \gamma(k_1) + \gamma(k_2)~,
\end{align}
is presented (see Fig.~\ref{fig:real}). 

We follow the helicity amplitude method with the conventions 
introduced by~\cite{Jegerlehner:2000wu,Kolodziej:1991pk}.
The Weyl representation for fermions is used where the Dirac matrices 
\begin{align}
\gamma^{\mu}= \left( 
\begin{array}{cc}
0 & \sigma^{\mu}_+ \\
\sigma^{\mu}_- & 0
\end{array} \right)~, \qquad \mu=0,1,2,3~,
\end{align}
are given in terms of the unit $2 \times 2$ matrix $I$ and the Pauli 
matrices $\sigma_i, i=1,2,3$, with $\sigma^{\mu}_{\pm}=(I,\pm\sigma_i)$.
The contraction of any four-vector $a^{\mu}$ with the $\gamma^{\mu}$ 
matrices has the form
\begin{align}
\ta{a} = a_{\mu} \gamma^{\mu} = \left( 
\begin{array}{cc}
0   & a^+ \\
a^- & 0
\end{array} \right)~,
\end{align}
where the $2 \times 2$ matrices $a^{\pm}$ are given by 
\begin{align}
a^{\pm} =  a^{\mu} \sigma_{\mu}^{\pm} &= \left( \begin{array}{cc}
a^0 \mp a^3      & \mp (a^1 - i a^2) \\
\mp (a^1 + i a^2) & a^0 \pm a^3
\end{array} \right)~.
\end{align}

The helicity spinors $u$ and $v$ for a particle and an antiparticle 
of four-momentum $p=(E, {\bf p})$ and helicity $\lambda/2=\pm 1/2$ 
are given by
\begin{align}
u(p,\lambda) &= \left( \begin{array}{c}
\sqrt{E-\lambda |{\bf p}|}\; \chi({\bf p},\lambda) \\
\sqrt{E+\lambda |{\bf p}|}\; \chi({\bf p},\lambda)
\end{array} \right) \equiv \left( \begin{array}{c}
u_I \\ u_{II} \end{array} \right)~, \non \\ 
v(p,\lambda) &= \left( \begin{array}{c}
-\lambda \sqrt{E+\lambda |{\bf p}|}\; \chi({\bf p},-\lambda) \\
\lambda \sqrt{E-\lambda |{\bf p}|}\; \chi({\bf p},-\lambda)
\end{array} \right) \equiv \left( \begin{array}{c}
v_I \\ v_{II} \end{array} \right)~.
\end{align}
The helicity eigenstates $\chi({\bf p},\lambda)$ can be expressed
in terms of the polar and azimuthal angles of the momentum
vector ${\bf p}$ as
\begin{align}
\chi({\bf p},+1) &= \left( \begin{array}{r}
\cos{(\theta/2)} \\ e^{i\phi} \sin{(\theta/2)}
\end{array} \right)~, \non \\ 
\chi({\bf p},-1) &= \left( \begin{array}{r}
-e^{-i\phi}\sin{(\theta/2)} \\ \cos{(\theta/2)}
\end{array} \right)~.
\end{align}
Finally, complex polarization vectors in the helicity basis are defined
for the real photons: 
\begin{align}
\varepsilon^{\mu}(k_i,\lambda_i=\mp) = \frac{1}{\sqrt{2}} \big( & 0,
\pm \cos \theta_i \cos \phi_i + i \sin \phi_i, \non \\ &
\pm \cos \theta_i \sin \phi_i - i \cos \phi_i,
\mp \sin{\theta_i} \big)~,
\end{align}  
with $i=1,2$.

The complete amplitude can be written in the following form
\begin{eqnarray}
 \kern-15pt M\left(\lambda_{e^+},\lambda_{e^-},\lambda_1,\lambda_2\right) =
 &&v_I^{\dagger}(\lambda_{e^+}) A\left(\lambda_1,\lambda_2\right)
 u_I(\lambda_{e^-}) 
 \nonumber \\
   + &&v_{II}^{\dagger}(\lambda_{e^+}) B\left(\lambda_1,\lambda_2\right)
 u_{II}(\lambda_{e^-})
 \ , 
\label{a1}
\end{eqnarray}
 where \(A\left(\lambda_1,\lambda_2\right)\) and 
 \(B\left(\lambda_1,\lambda_2\right)\) are $2 \times 2$ matrices defined in 
 the appendix with the matrix elements $A_{i,j}(\lambda_1,\lambda_2)$ and
 $B_{i,j}(\lambda_1,\lambda_2)$. It simplifies even further, when calculated 
 in the electron--positron cms frame (the $z$-axis was chosen 
 along the positron momentum):
 \begin{eqnarray}
 M\left(+,+,\lambda_1,\lambda_2\right) &=& 
 m_e \left(A_{2,2}\left(\lambda_1,\lambda_2\right)
 -B_{2,2}\left(\lambda_1,\lambda_2\right)\right)~, \nonumber \\
 M\left(-,-,\lambda_1,\lambda_2\right) &=& 
 m_e \left(A_{1,1}\left(\lambda_1,\lambda_2\right)
 -B_{1,1}\left(\lambda_1,\lambda_2\right)\right)~, \nonumber \\
  M\left(+,-,\lambda_1,\lambda_2\right) &=&
 -\left(E+|{\bf p}|\right)A_{2,1}\left(\lambda_1,\lambda_2\right)\nonumber \\
&&+\frac{m_e^2}{E+|{\bf p}|}B_{2,1}\left(\lambda_1,\lambda_2\right)~,
\nonumber \\
  M\left(-,+,\lambda_1,\lambda_2\right) &=&
 \left(E+|{\bf p}|\right)B_{1,2}\left(\lambda_1,\lambda_2\right)\nonumber \\
 &&-\frac{m_e^2}{E+|{\bf p}|}A_{1,2}\left(\lambda_1,\lambda_2\right)~.
\label{a2}
\end{eqnarray}

From the explicit form of the matrices \(A\left(\lambda_1,\lambda_2\right)\) 
and \(B\left(\lambda_1,\lambda_2\right)\),
it is clear that some factors appear repeatedly in different amplitudes. 
In order to speed up the numerical computation, the amplitudes 
are decomposed into these factors, which are used as building 
blocks for all the diagrams. Then, a polarized matrix element is 
calculated numerically for a given set of external particle momenta
in a fixed reference frame, e.g. the cms of the 
initial particles where the initial momenta are parallel to the $z$ axis. 
The result is squared and the sum over polarizations is performed. 

As a test, the square of the matrix element averaged over initial 
particle polarization has also been calculated using the standard 
trace technique and tested numerically against the helicity method result. 
Perfect agreement between the two approaches is found.
Both matrix elements vanish if the photon polarization vectors are replaced
by their four-momenta, and thus are tested for gauge invariance.

\section{Monte Carlo simulation}

On the basis of these results a Monte Carlo generator, called PHOKHARA,
has been built to simulate the production 
of two charged pions together with one or two hard photons; it includes 
virtual and soft photon corrections to the emission of one single real 
photon. It supersedes the previous versions of the 
EVA~\cite{Binner:1999bt} Monte Carlo generator.
The program exhibits a modular structure, which preserves the 
factorization of the initial state QED corrections. The simulation 
of other exclusive hadronic channels can therefore be easily included 
with the simple replacement of the current(s) of the existing modes, 
and the corresponding multiparticle hadronic phase space. 
The simulation of the four-pion channel~\cite{Czyz:2000wh} will be 
incorporated soon, as well as other multihadron final states. 

The program provides predictions either at LO or at NLO. In the 
former case only single photon events are generated. 
In the latter, both events with one or two photons are 
generated at random. For simplicity, FSR is not considered 
in the new generator, which however can be estimated from 
EVA~\cite{Binner:1999bt}. 

Single photon events are generated following the same procedure 
as in EVA. The generation of two-photon events proceeds 
as follows. First, polar and azimuthal angles of the two photons 
are generated. One of the polar angles is generated 
within the given angular cuts, the other is generated unbounded. 
In this way, the photon angular cuts are automatically fulfilled
and a higher generation efficiency is achieved. Both photons are 
nevertheless symmetrically generated. Then, the hadronic invariant 
mass $Q^2$ is generated following the resonant distribution of the 
hadronic current, its maximum being determined by~\eq{dpslimits} 
in Appendix~\ref{app:phasespace},
where one of the photon energies is set to the soft photon cutoff
and the other is set to the minimal detection photon energy 
$E_{\gamma}^{min}$. Next, the photon energies are generated. 
If only one of the photons passes the angular cuts, its energy 
is forced to be larger than $E_{\gamma}^{min}$. The other 
photon energy is calculated according to~\eq{dpslimits}.
Otherwise, if both photons pass the angular cuts the minimal energy
of one of them is fixed to $w$, with equal probability for both photons,  
and the other is calculated through~\eq{dpslimits}.
Finally, the hadron momenta are generated in the $Q^2$ rest frame
and then boosted to the $e^+ e^-$ cms. 
Further angular cuts or other kinds of constraints are imposed
after generation. For more details, see Appendix~\ref{app:phasespace}.

\section{Tests and results}

\subsection{Gauss integration versus MC}
 
 To test the technical precision of the single photon generation, 
 a FORTRAN program was written, which performs the two-dimensional 
 integral that remains after the integration over the pion angles, and 
 the photon azimuthal angle has been performed analytically with the 
 help of the relation
\begin{eqnarray}
 \int \ J^{2\pi}_\mu (J^{2\pi}_\nu)^*  \ \ d \Phi_2(Q;q_1,q_2) = 
 \nonumber \\
 \frac{e^2}{6\pi} \left(Q_{\mu}Q_{\nu}-g_{\mu\nu}Q^2\right) \ R(Q^2) \ \ .
\label{rr}
\end{eqnarray}
 
 For the remaining numerical integration, the integration region was 
 sliced into an appropriate number of subintervals (typically 100
 to 200), and 8-point Gauss quadrature was used in each of the
 subintervals. This leads to a relative accuracy of 10\(^{-10}\). 

\begin{figure}
\begin{center}
\epsfig{file=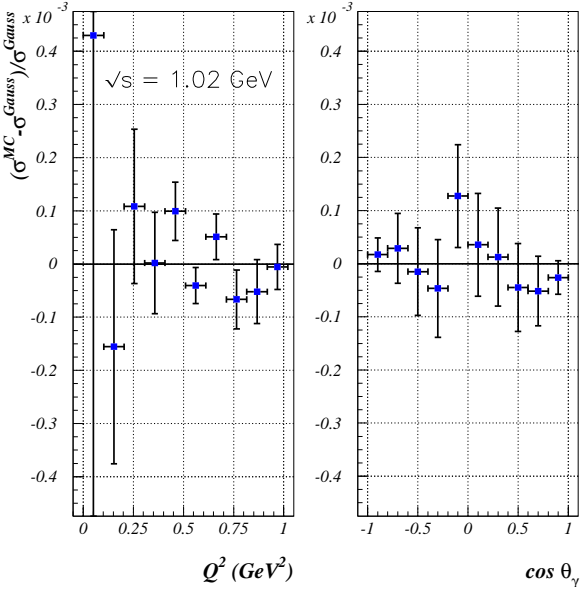,width=8.5cm} 
\end{center}
\caption{Relative difference between the single photon cross section obtained 
 from the Monte Carlo generator \(\sigma^{MC}\) and Gauss quadrature
 \(\sigma^{Gauss}\) (see text for details).}
\label{fig:gau2}
\end{figure}

 The test was performed for a photon polar angle \(\theta_{\gamma}\)
 between 10\(^{\circ}\) and 170\(^{\circ}\) and for
 \( 4m_{\pi}^2 < Q^2 < 2E_{cm}(E_{cm}-E_{\gamma}^{min} )\)
 with \(E_{cm} = 1.02 \) GeV and \(E_{\gamma}^{min} = 10\) MeV.
 To study contributions from different ranges of \(\theta_{\gamma}\) 
 and \(Q^2\) the above-mentioned intervals of \(Q^2\) and 
 \(\cos(\theta_{\gamma})\)  were divided into ten equally spaced parts.
 The integrals were performed first within the whole range 
 of \(\theta_{\gamma}\), and ten subintervals 
 in \(Q^2\) separately and subsequently within the whole range
 of \(Q^2\) and ten subintervals of \(\cos(\theta_{\gamma})\).
 From Fig.\ref{fig:gau2} it is clear that a technical
 precision of the order of 10\(^{-4}\) was achieved. The error bars indicate
 one standard deviation of the Monte Carlo generator, which performs 
 the five-dimensional integral.

\subsection{Soft photon cutoff independence}

\begin{figure}
\begin{center}
\epsfig{file=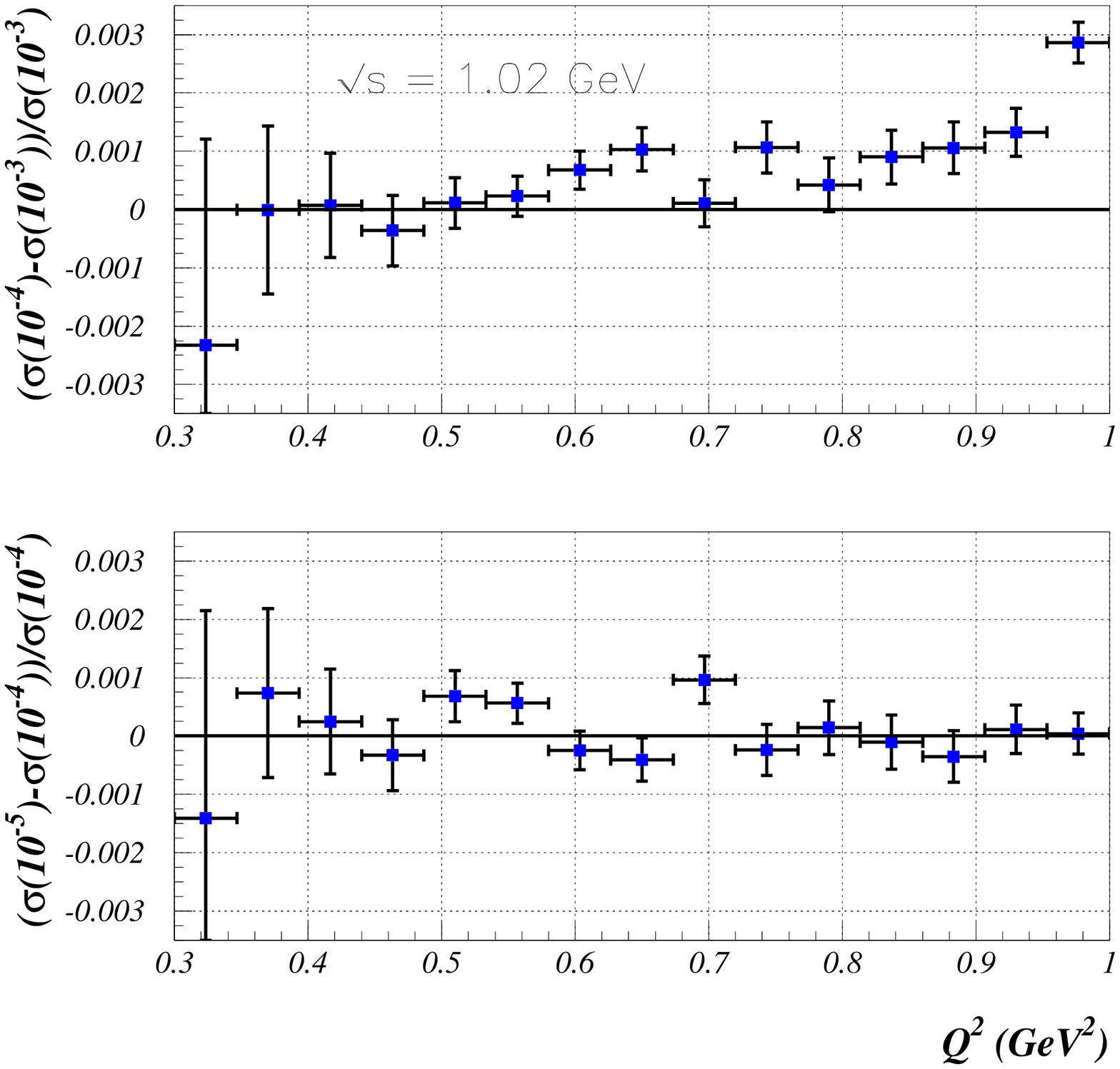,width=8.5cm}
\end{center}
\caption{Comparison of the $Q^2$ differential distribution for different 
values of the soft photon cutoff: $w = 10^{-3}$, $10^{-4}$ and $10^{-5}$,
at $\sqrt{s}=1.02$~GeV.}
\label{fig:epstest}
\end{figure}

\begin{figure}
\begin{center}
\epsfig{file=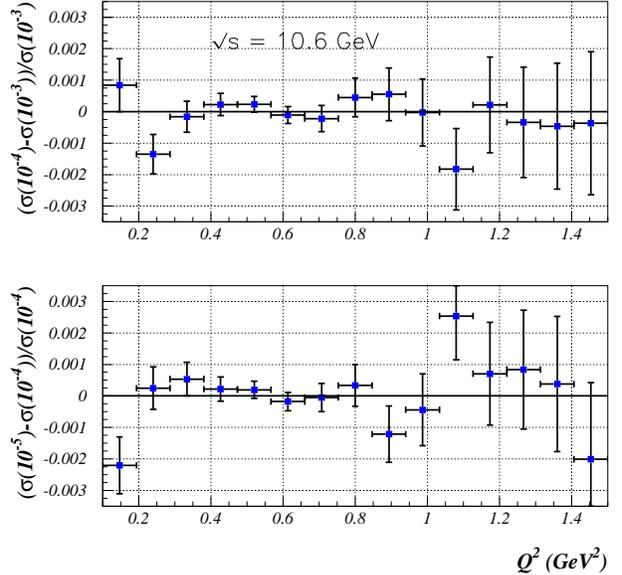,width=8.5cm}
\end{center}
\caption{Comparison of the $Q^2$ differential distribution for different 
values of the soft photon cutoff: $w = 10^{-3}$, $10^{-4}$ and $10^{-5}$,
at $\sqrt{s}=10.6$~GeV.}
\label{fig:epstestb}
\end{figure}

The full NLO calculation consists of two complementary contributions,
the virtual and soft corrections presented in section~\ref{sec:virtsoft}
and the hard correction described in section~\ref{sec:two}.
The former depends logarithmically on the soft photon cutoff $w$, 
see~\eq{nloleptonic}. The second, after numerical integration 
in phase space exhibits the same behaviour, so that their sum 
must be independent of $w$. 

However, a particular value of $w$ has to be chosen for the generation.
To be valid, the soft photon approximation requires $w$ to be small. 
On the other hand a very small value of $w$ could even produce 
unphysical negative weights for the generated events. The particular 
value of $w$ chosen to perform the Monte Carlo generation should 
therefore arise from a compromise between these two conditions.

In this section we show that the result from the generator is indeed
independent from the soft photon cutoff $w$, within the 
error of the numerical integration. Furthermore, we try to 
determine the value of $w$ that optimizes the event generation.

\begin{table}
\caption{Kinematical cuts applied at different cms energies:
minimal energy of the tagged photon ($E_{\gamma}$), angular 
cuts on the tagged photon ($\theta_{\gamma}$) and the pions ($\theta_{\pi}$),
and minimal invariant mass of the hadrons plus the tagged photon
($M^2_{\pi^+\pi^-\gamma}$)}
\label{tab:cuts}
\begin{center}
\begin{tabular}{cccc}
                   & $\sqrt{s}=$1.02~GeV & 4~GeV & 10.6~GeV \\ \hline  
$E_{\gamma}^{min}$ (GeV) & $0.01$ & $0.1$ & $1$  \\
$\theta_{\gamma}$ (degrees)  & $[5,21]$   & $[10,170]$ & $[25,155]$ \\
$\theta_{\pi}$ (degrees)     & $[55,125]$ & $[20,160]$ & $[30,150]$ \\
$M^2_{\pi^+\pi^-\gamma}$ (GeV$^2$) & $0.9$ & $12$ & $90$ \\ 
\hline
\end{tabular}
\end{center}
\end{table}

The tests have been performed for different cms energies, 
from 1 to 10~GeV, corresponding to the energies of DA$\Phi$NE, CLEO-C and 
$B$-meson factories. Kinematical cuts have been applied as listed in 
Table~\ref{tab:cuts} and will be used for the rest of the paper.
They are related to those of the experiments for which we present our 
predictions and at the same time allow final state radiation to be controlled, 
one of the key points of the radiative return method. A minimal energy 
$E_{\gamma}^{min}$ is required for the tagged photon. 
Different cuts are chosen for the polar angle of the tagged photon 
$\theta_{\gamma}$ at different energies. At low energies, the pions 
are constrained to be well separated from the photons to suppress 
final state radiation. At high energies, the observed photon and the 
pions are mainly produced back to back; the suppression 
of the final state radiation is therefore naturally accomplished. 
Furthermore, a minimal invariant mass of the hadronic system plus the 
tagged photon, $M^2_{\pi^+\pi^-\gamma}$, is required. 
The reason for this last kinematical cut will be discussed later.
When events with two photons are simulated we require at least one 
of the photons to pass the cuts.

\begin{table}
\caption{Total cross section (nb) for the process 
$e^+ e^- \rightarrow \pi^+ \pi^- \gamma$ at NLO for different values 
of the soft photon cutoff. Only initial state radiation.
Cuts from Table~\ref{tab:cuts}.}
\label{tab:epstest}
\begin{center}
\begin{tabular}{cccc}
$w$ & $\sqrt{s}=$1.02~GeV & 4~GeV & 10.6~GeV \\ \hline 
$10^{-3}$ & 2.0324 (4) & 0.12524 (5) & 0.010564 (4)\\
$10^{-4}$ & 2.0332 (5) & 0.12526 (5) & 0.010565 (4)\\
$10^{-5}$ & 2.0333 (5) & 0.12527 (5) & 0.010565 (5)\\ \hline
\end{tabular}
\end{center}
\end{table}

Table~\ref{tab:epstest} presents the total cross section calculated 
for several values of the soft photon cutoff at three different cms 
energies for the kinematical cuts from Table~\ref{tab:cuts}. 
The excellent agreement confirms the $w$-independence of the result. 

Figures~\ref{fig:epstest} and~\ref{fig:epstestb} show the $w$-independence 
of the $Q^2$ distribution at $\sqrt{s}=1.02$~GeV and $10.6$~GeV cms 
energies respectively. Even if the analysis of the total cross 
section (Table~\ref{tab:epstest}) suggests that the choice of $w=10^{-3}$
is as good as  $w=10^{-4}$, small differences in the differential
cross section are found for high values of $Q^2$. Similar comparisons 
between the differential cross section calculated for $w=10^{-4}$ and 
$w=10^{-5}$ show a perfect agreement within the statistical errors, well 
below the 0.1\% level. As a result we chose $w=10^{-4}$ for the soft 
photon cutoff.

\subsection{Comparison of NLO results with the structure function
approximation and an estimation of the theoretical error}

\begin{table}
\caption{Total cross section (nb) for the process 
$e^+ e^- \rightarrow \pi^+ \pi^- \gamma$ at LO, NLO(1) and in the 
collinear approximation via structure functions (SF) with the cuts from 
Table~\ref{tab:cuts}. Only initial state radiation. NLO(2) gives 
the NLO result with the same cuts as NLO(1) but for the hadron--photon 
invariant mass unbounded.}
\label{tab:xsec}
\begin{center}
\begin{tabular}{lccc}
  & $\sqrt{s}=$1.02~GeV & 4~GeV & 10.6~GeV \\ \hline  
Born     & 2.1361 (4) & 0.12979 (3) & 0.011350 (3) \\
SF       & 2.0192 (4) & 0.12439 (5) & 0.010526 (3) \\
NLO (1)  & 2.0332 (5) & 0.12526 (5) & 0.010565 (4) \\ 
NLO (2)  & 2.4126 (7) & 0.14891 (9) & 0.012158 (9) \\ 
\hline
\end{tabular}
\end{center}
\end{table}

\begin{table}
\caption{Total cross section (nb) for the process 
$e^+ e^- \rightarrow \pi^+ \pi^- \gamma$ at $\sqrt{s}=$1.02~GeV in NLO 
and in the collinear approximation (SF) as a function of the cut on the 
invariant mass of the hadron + tagged photon $M^2_{\pi^+ \pi^- \gamma}$.
Only initial state radiation. Minimal energy of the tagged photon and 
angular cuts from Table~\ref{tab:cuts}.}
\label{tab:invariantcut}
\begin{center}
\begin{tabular}{ccc}
$M^2_{\pi^+ \pi^- \gamma}$ (GeV$^2$) & SF & NLO \\ \hline  
0.1  & 2.4127(18) & 2.4132(8)\\
0.2  & 2.4126(18) & 2.4131(8)\\
0.3  & 2.4124(18) & 2.4127(8)\\
0.4  & 2.4098(18) & 2.4096(8)\\
0.5  & 2.3949(18) & 2.3953(8)\\
0.6  & 2.3425(16) & 2.3455(8)\\
0.7  & 2.2449(11) & 2.2543(8)\\
0.8  & 2.1387(9) & 2.1533(8)\\
0.9  & 2.0198(8) & 2.0334(8)\\
0.95 & 1.9437(8) & 1.9522(8)\\
0.99 & 1.8573(8) & 1.8559(8)\\
\hline
\end{tabular}
\end{center}
\end{table}
\begin{figure}
\begin{center}
\epsfig{file=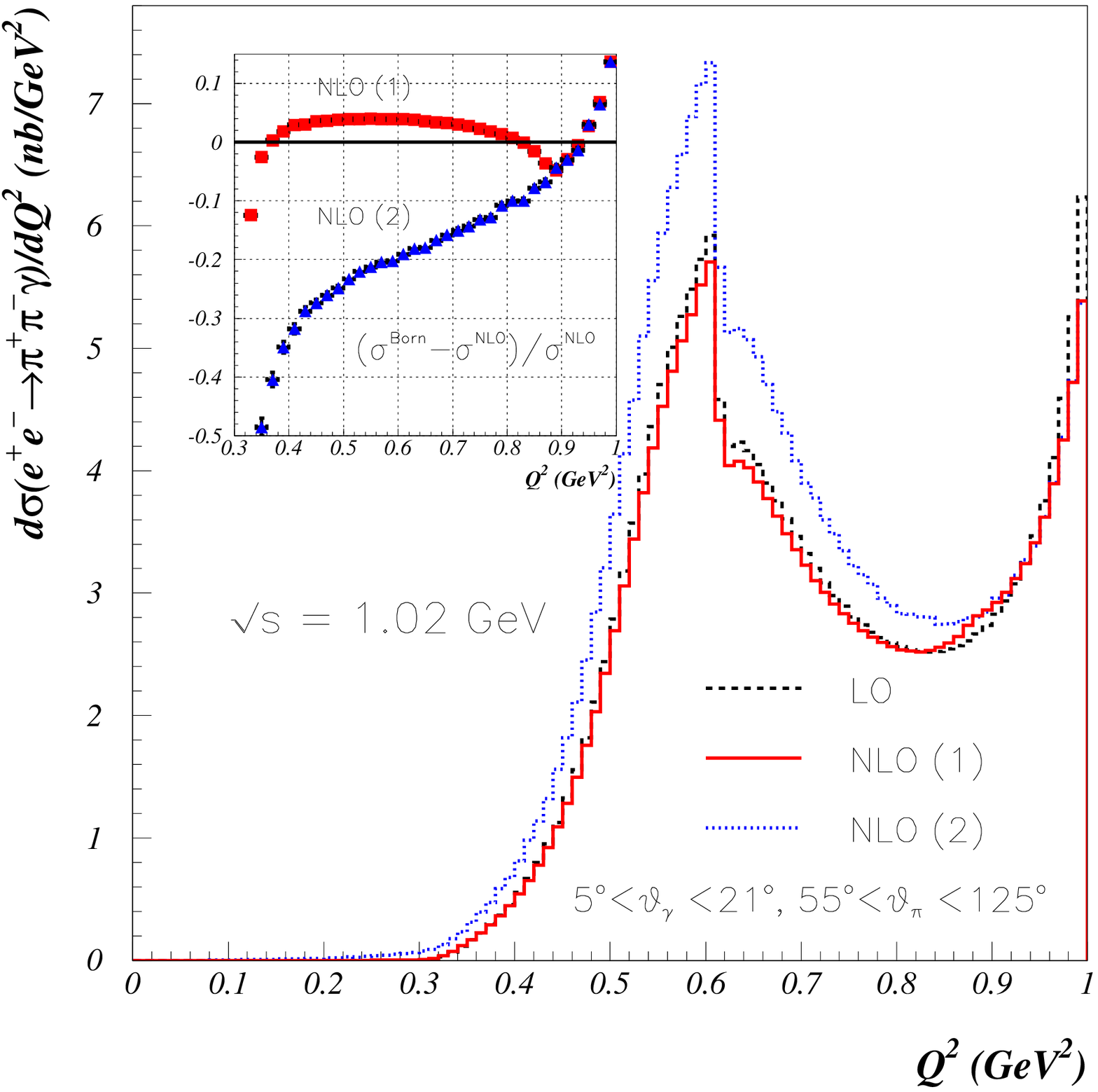,width=8.5cm} 
\end{center}
\caption{Differential cross section for the process 
$e^+ e^- \rightarrow \pi^+ \pi^- \gamma$ at NLO for $\sqrt{s}=1.02$~GeV. 
Only initial state radiation. The relative size of the correction to 
the LO result (dashed) is shown in the small inset. The cuts are: 
$5^\circ<\theta_{\gamma}<21^\circ$, $55^\circ<\theta_{\pi}<125^\circ$,
the energy of the tagged photon $E_{\gamma}>0.01$~GeV and the 
invariant mass of the detected particles in the final state 
$M^2_{\pi^+ \pi^- \gamma}>0.9$~GeV$^2$ for NLO(1) (solid).
NLO(2) (dotted) obtained without the last cut.}
\label{fig:nlo1gev}
\end{figure}

\begin{figure}
\begin{center}
\epsfig{file=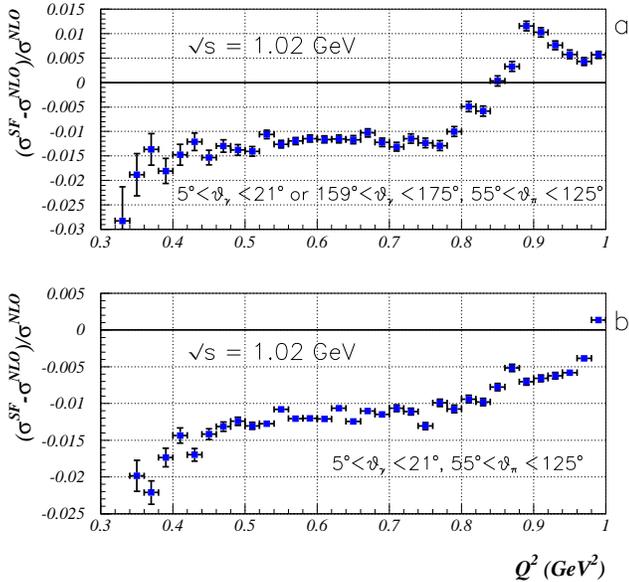,width=8.5cm} 
\end{center}
\caption{Comparison between the collinear approximation
by structure functions and the fixed order NLO result. 
Cuts from Table~\ref{tab:cuts} for the lower figure.
In the upper figure, same cuts as below, with the addition of
the symmetric photon configuration $159^\circ<\theta_{\gamma}<175^\circ$.}
\label{fig:collnlo}
\end{figure}

\begin{figure}
\begin{center}
\epsfig{file=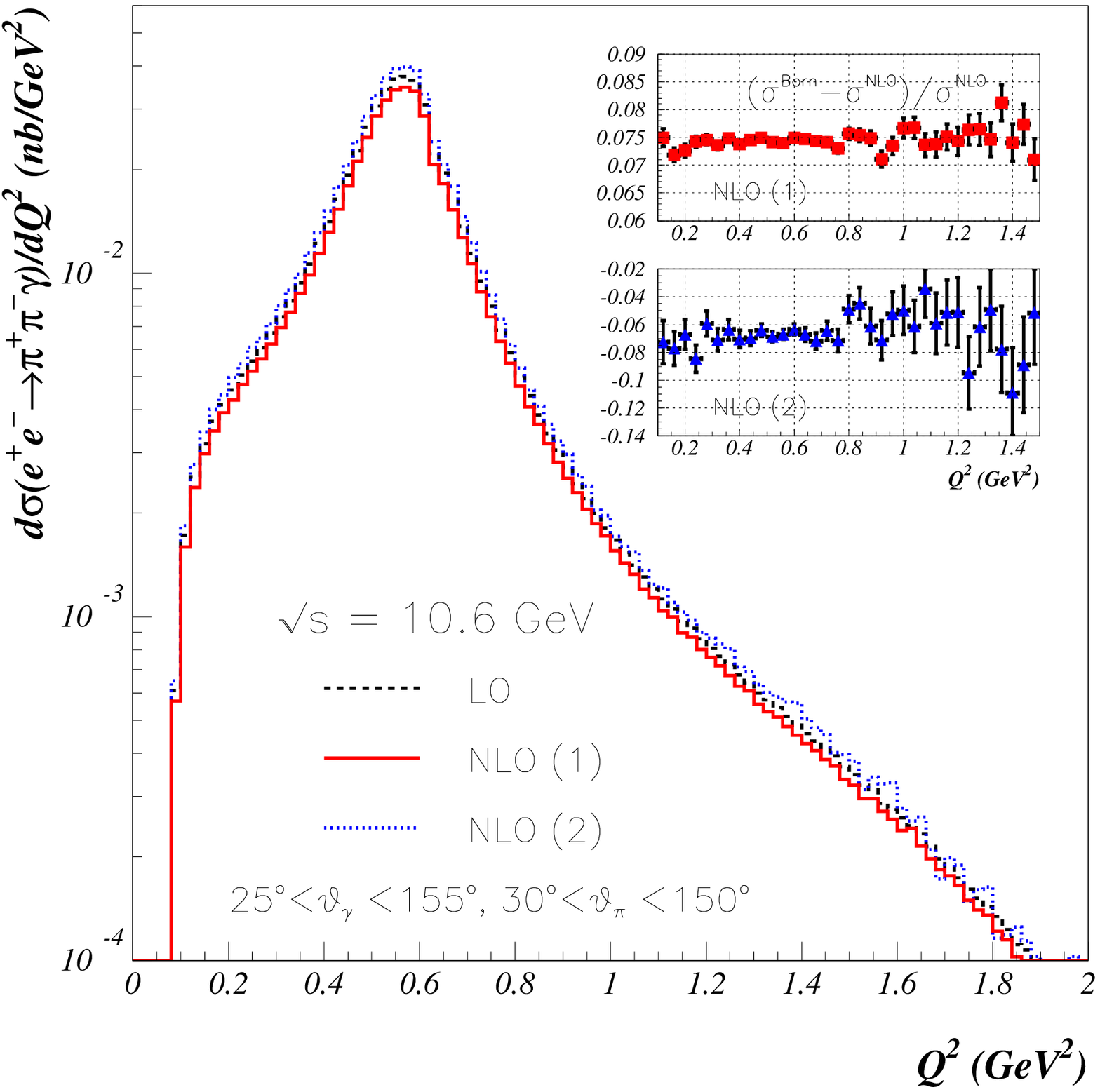,width=8.5cm} 
\end{center}
\caption{Differential cross section for the process 
$e^+ e^- \rightarrow \pi^+ \pi^- \gamma$ at NLO for $\sqrt{s}=10.6$~GeV. 
Only initial state radiation. The relative size of the correction to 
the LO result (dashed) is shown in the small inset. The cuts are: 
$25^\circ<\theta_{\gamma}<155^\circ$, $30^\circ<\theta_{\pi}<150^\circ$,
the energy of the tagged photon $E_{\gamma}>1$~GeV and the 
invariant mass of the detected particles in the final state 
$M^2_{\pi^+ \pi^- \gamma}>90$~GeV$^2$ for NLO(1) (solid).
NLO(2) (dotted) obtained without the last cut.}
\label{fig:nlo10gev}
\end{figure}

\begin{figure}
\begin{center}
\epsfig{file=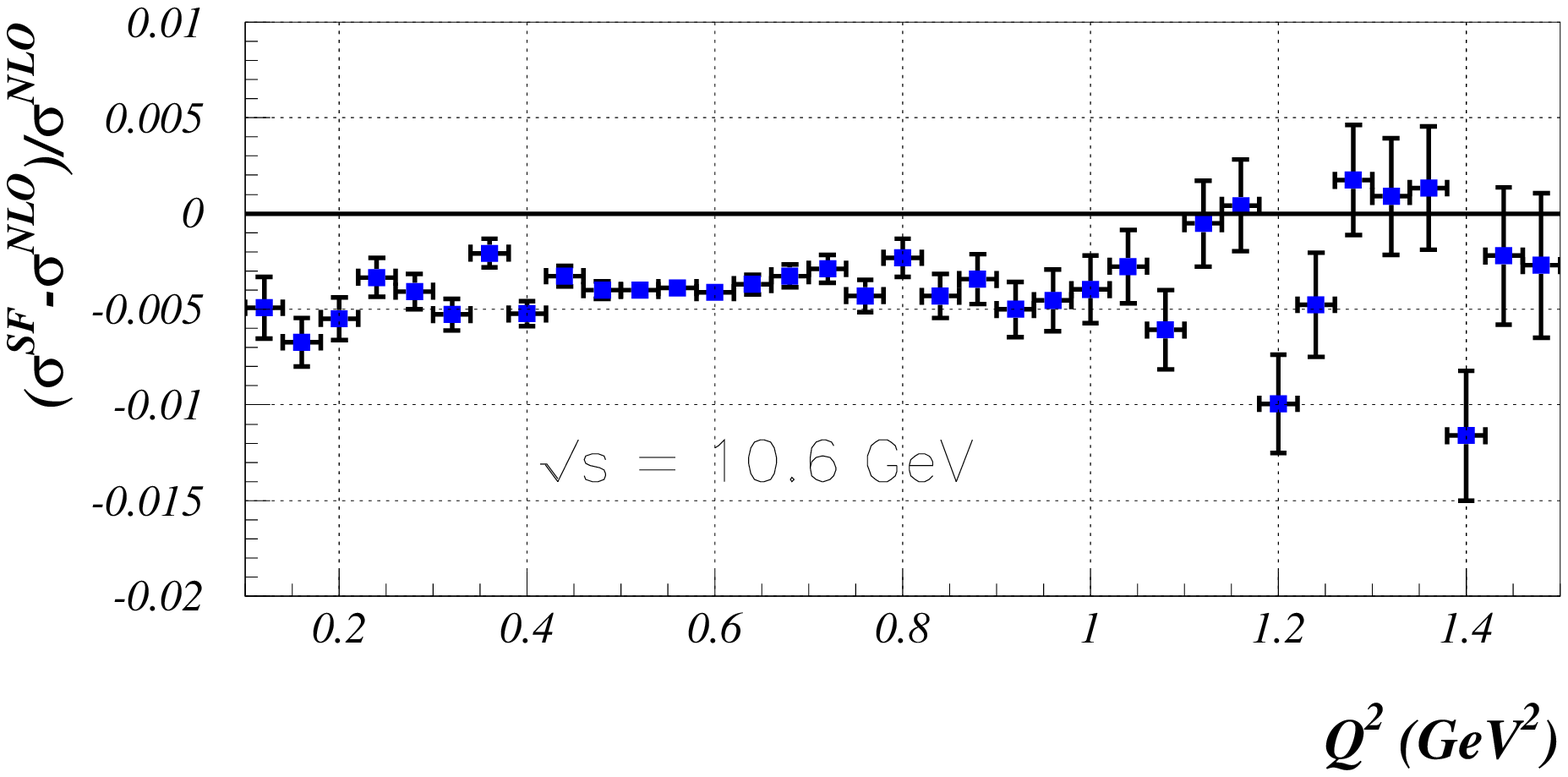,width=8.5cm} 
\end{center}
\caption{Comparison between the collinear approximation
by structure functions and the fixed order NLO result. 
Cuts from Table~\ref{tab:cuts}.}
\label{fig:collnlob}
\end{figure}

\begin{figure}
\begin{center}
\epsfig{file=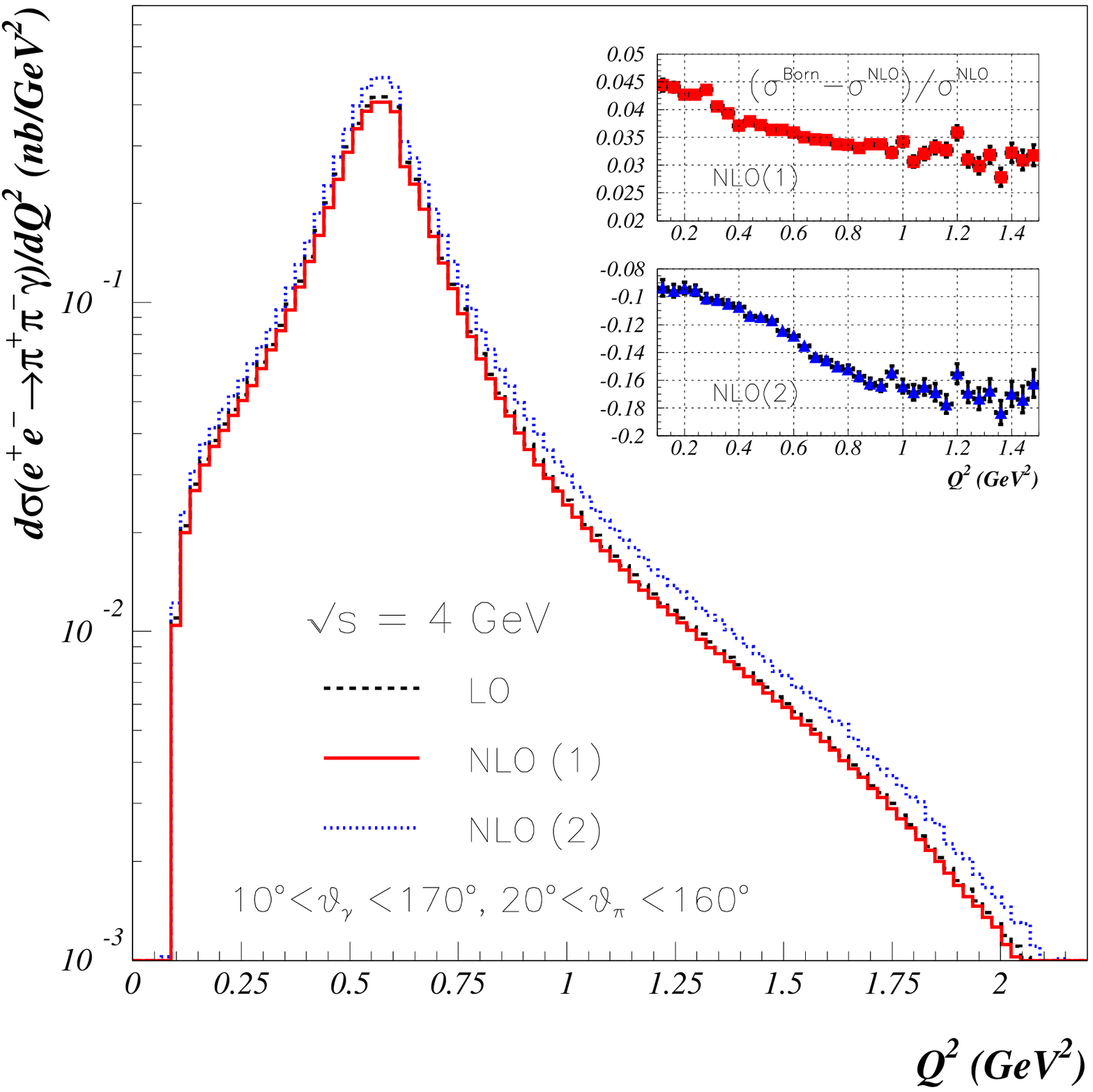,width=8.5cm} 
\end{center}
\caption{Differential cross section for the process 
$e^+ e^- \rightarrow \pi^+ \pi^- \gamma$ at NLO for $\sqrt{s}=4$~GeV.  
Only initial state radiation. The relative size of the correction to 
the LO result (dashed) is shown in the small inset. The cuts are: 
$10^\circ<\theta_{\gamma}<170^\circ$, $20^\circ<\theta_{\pi}<160^\circ$, 
the energy of the tagged photon $E_{\gamma}>0.1$~GeV and the 
invariant mass of the detected particles in the final state 
$M^2_{\pi^+ \pi^- \gamma}>12$~GeV$^2$ for NLO(1) (solid).
NLO(2) (dotted) obtained without the last cut.}
\label{fig:nlo4gev}
\end{figure}

\begin{figure}
\begin{center}
\epsfig{file=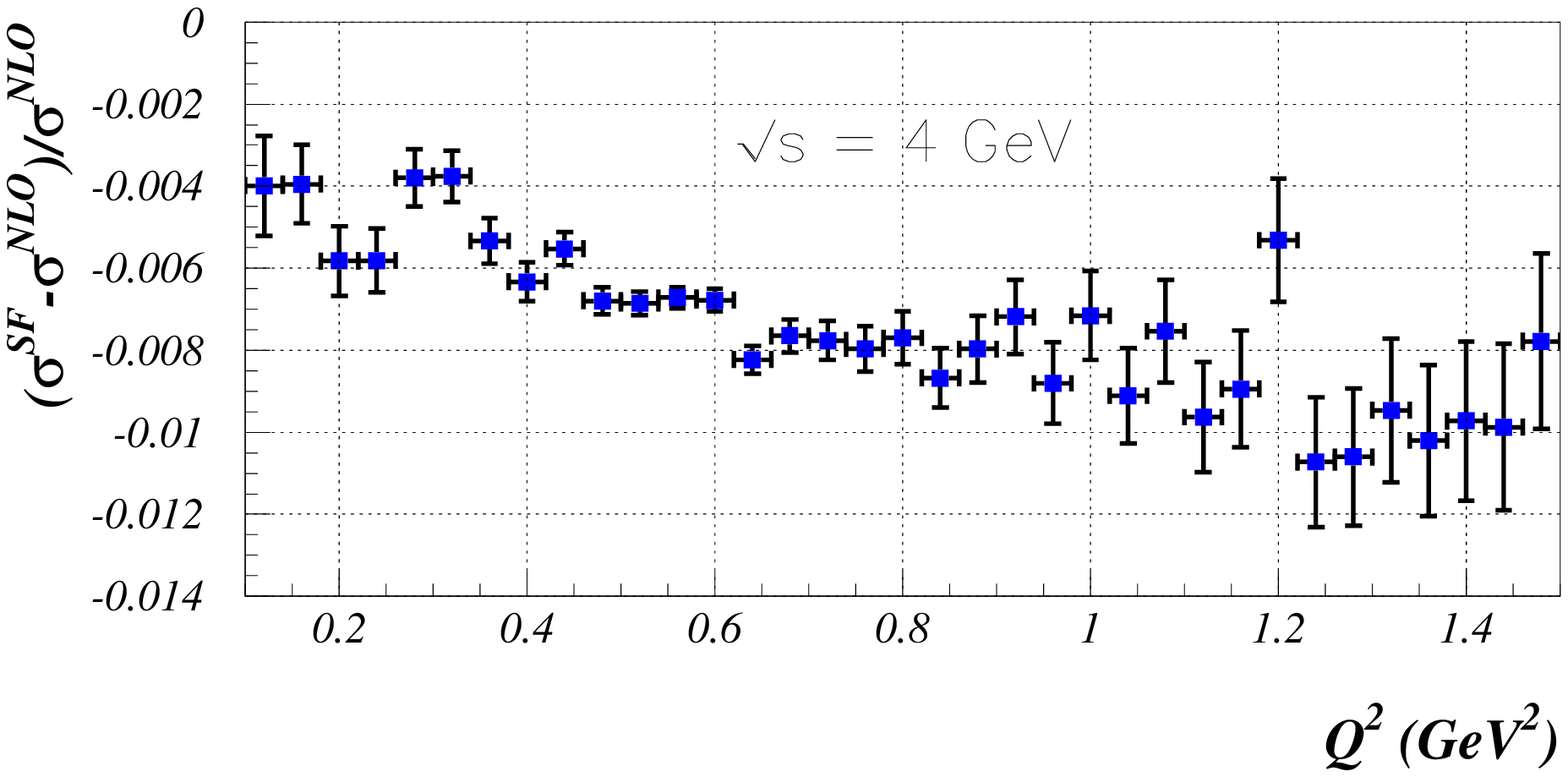,width=8.5cm} 
\end{center}
\caption{Comparison between the collinear approximation
by structure functions and the fixed order NLO result. 
Cuts from Table~\ref{tab:cuts}.}
\label{fig:collnloc}
\end{figure}

The original and default version of EVA~\cite{Binner:1999bt}, simulating 
the process $e^+ e^- \rightarrow \pi^+ \pi^- \gamma$ at LO, allowed for 
additional initial state radiation of soft and collinear photons by the 
structure function method~\cite{structure,Caffo:1997yy}. By convoluting 
the Born cross section with a given SF distribution,
the soft photons are resummed to all orders in perturbation theory 
and large logarithms of collinear origin, $L=\log(s/m_e^2)$, are taken 
into account up to two-loop approximation. 
The NLO result, being a complete one-loop result, contains these logarithms 
in order $\alpha$ and the additional subleading terms, which of course
are not taken into account within the SF method. 
The subleading terms from virtual and soft corrections were calculated 
in~\cite{Rodrigo:2001jr} and are included in the present NLO generator. 

In the SF approach, the additional emission of collinear 
photons reduces the effective cms energy of the collision.
In~\cite{Binner:1999bt}, a minimal invariant mass of the observed 
particles, hadrons plus tagged photon, was required in order to reduce 
the kinematic distortion of the events. To perform the comparison 
between EVA and the present program a similar cut is introduced in the 
NLO calculation. For one-photon events the invariant mass of the hadrons 
and the emitted photon is equal to the total cms energy
and the requirement is trivially fulfilled. However, when two-photon 
events are generated, energy and production angles of both photons
are compared with the cuts listed in Table~\ref{tab:cuts}. If one of 
the photons fulfils both requirements, its common invariant mass 
with the hadrons is calculated. In other words, we require at least 
one of the photons to pass all the cuts, including the one on its common 
invariant mass with the hadrons. The probability of both photons passing 
all the cuts becomes very small when this invariant mass cut is close to 
the total cms energy. 

Table~\ref{tab:xsec} gives the total cross section calculated at LO 
and NLO for the previous kinematical cuts. The soft photon cutoff is 
fixed at $w=10^{-4}$. For comparison, the total cross section 
derived from EVA with emission of collinear photons by structure 
functions~\cite{Caffo:1997yy,structure} is presented. Two NLO 
predictions are shown. The first one, NLO(1), which can be compared 
with the SF result, includes the cut on the invariant mass of the hadrons 
plus photon. The second one, NLO(2), is obtained without this cut. 
The results of EVA and those denoted NLO(1) for the total cross 
section are in agreement to within $0.7\%$. Both of them are 
clearly sensitive to the cut on $M^2_{\pi^+ \pi^- \gamma}$.
This cut dependence is displayed in Table~\ref{tab:invariantcut}.
Remarkably enough, the typical difference between the results of the 
two programs is clearly less than $0.5\%$ for most of the entries.

Figure~\ref{fig:nlo1gev} presents our NLO predictions 
for the differential cross section of the process 
$e^+ e^- \rightarrow \pi^+ \pi^- \gamma$ as a function of 
$Q^2$, the invariant mass of the hadronic system, at DA$\Phi$NE energies, 
$\sqrt{s} = 1.02$~GeV, with the same kinematical conditions as before. 
For comparison, also the LO prediction (without collinear emission)
is shown, as well as the relative correction.
Notice that although the NLO(1) prediction for the total cross section 
differs from the LO result by roughly $5\%$ (see Table~\ref{tab:xsec}), 
the $Q^2$ distribution shows corrections up to $\pm 15\%$ at very low 
or high values of $Q^2$. 

As already shown in Table~\ref{tab:xsec} for the total cross section, 
the invariant mass of the hadron + tagged photon system of a sizeable 
fraction of events lies below the cut of $0.9$~GeV$^2$. These are events 
where a second hard photon must be present. Including these events leads 
to the distribution denoted NLO(2) in Fig.~\ref{fig:nlo1gev}.

In contrast to the typically $5\%$ difference between LO and NLO(1)
predictions, only a $0.6\%$ difference is found between the NLO(1) 
prediction and the SF collinear approximation as implemented in EVA, 
this difference being higher at small $Q^2$ 
and lower at high $Q^2$, see Fig.~\ref{fig:collnlo}. 
As one can see from  Fig.~\ref{fig:collnlo}, the size and sign of the 
NLO corrections do depend on the particular choice of the experimental
cuts. Hence only using a Monte Carlo event generator one can realistically 
compare theoretical predictions with experiment and extract \(R(s)\) 
from the data. The difference between Figs.~\ref{fig:collnlo}a 
and~\ref{fig:collnlo}b 
arises from the (small) subset of events with two photons, which 
both fulfil the angular and energy cuts and thus enter only once in 
the sample of Fig.~\ref{fig:collnlo}b.

Results at $10.6$~GeV cms energy are presented in 
Figs.~\ref{fig:nlo10gev} and~\ref{fig:collnlob}. In this case a 
NLO(1) correction to the LO result of $7.5\%$ is found, almost 
independent of $Q^2$, the difference between the NLO(1) prediction 
and the SF collinear approximation being smaller 
than $0.5\%$ and also almost independent of $Q^2$.

Finally, predictions for $\sqrt{s}=4$~GeV are presented in 
Figs.~\ref{fig:nlo4gev} and~\ref{fig:collnloc}.

To estimate the systematic uncertainty of the program, we observe that
leading logarithmic two-loop ${\cal O}(\alpha^2)$ corrections and the 
associate real emission are not included. The difference between the LO 
and the NLO(1) results was expected to be of the order of  
$\frac{3}{2} (\alpha/\pi) \log(s/m_e^2) \approx 5\%$ at $\sqrt{s}=1$~GeV.
Indeed one observes (see Fig.~\ref{fig:nlo1gev}) values typically of this
magnitude with maximal deviations close to $10\%$.
From na\"{\i}ve exponentiation one expects 
$\frac{1}{2} ( \frac{3}{2} (\alpha/\pi) \log(s/m_e^2) )^2 \approx 0.1$--$0.2\%$
for the leading photonic next-to-next-to-leading order (NNLO)
terms, which are ignored in the present program. 

Another type of $\alpha^2$ corrections originates from fermion 
(dominantly electron) loop insertions in the one-loop virtual 
corrections considered in \cite{Rodrigo:2001jr}.
These are conveniently combined with the emission of (mainly soft and
collinear) real fer\-mions, i.e. with the process 
$e^+ e^- \rightarrow \gamma \gamma^* f \bar{f}$, where the collinear
$f \bar{f}$ pair is mostly within the beam pipe and thus undetected. 
Individually these are corrections of order $(\alpha/\pi)^2 (\log(s/m_e^2))^3$.
When combined, the $\log^3(s/m_e^2)$ terms cancel and the 
remaining $\frac{1}{4}(\alpha/\pi)^2 (\log(s/m_e^2))^2$ term 
can be largely absorbed by choosing a running
coupling $\alpha$(1~GeV) in the NLO terms and thus amount to less
than $0.1\%$. Adding these contributions linearly, one estimates a
$0.3\%$ uncertainty~\footnote{Note that the reaction 
$e^+ e^- \rightarrow \gamma^* f \bar{f}$ leads to significantly 
larger effects~\cite{Kniehl:1988id,Hoefer:2001mx}. However, this
process does not contribute to events with a tagged photon.}.

Multiphoton emission is not included in the present program. In 
an inclusive treatment and for tight cuts on the photon energy,
this can in principle be included through exponentiation.
For the cuts on $M^2_{\pi^+ \pi^- \gamma}$ of $0.9$~GeV$^2$
proposed originally in~\cite{Binner:1999bt}, and adopted through most 
of this paper, the difference between the exponentiated form and the fixed 
order treatment (see e.g. eqs. 17 and 19 of~\cite{Rodrigo:2001jr}) amounts 
to roughly $0.7\%$, with smaller values for less restrictive cuts. 
For a cut of $M^2_{\pi^+ \pi^- \gamma}$ at $0.8$~GeV$^2$
or even $0.7$~GeV$^2$, which seems preferable from these considerations,
we expect a difference of $0.4\%$ and even below $0.3\%$ for the second
choice. In total we therefore estimate a systematic uncertainty from ISR 
of around $0.5\%$ in the total cross section, once loose cuts on 
$M^2_{\pi^+ \pi^- \gamma}$ are adopted. 
FSR can be controlled by suitable cuts or corrected with the 
Monte Carlo program EVA.


\subsection{Muon pair production}

\begin{figure}
\begin{center}
\epsfig{file=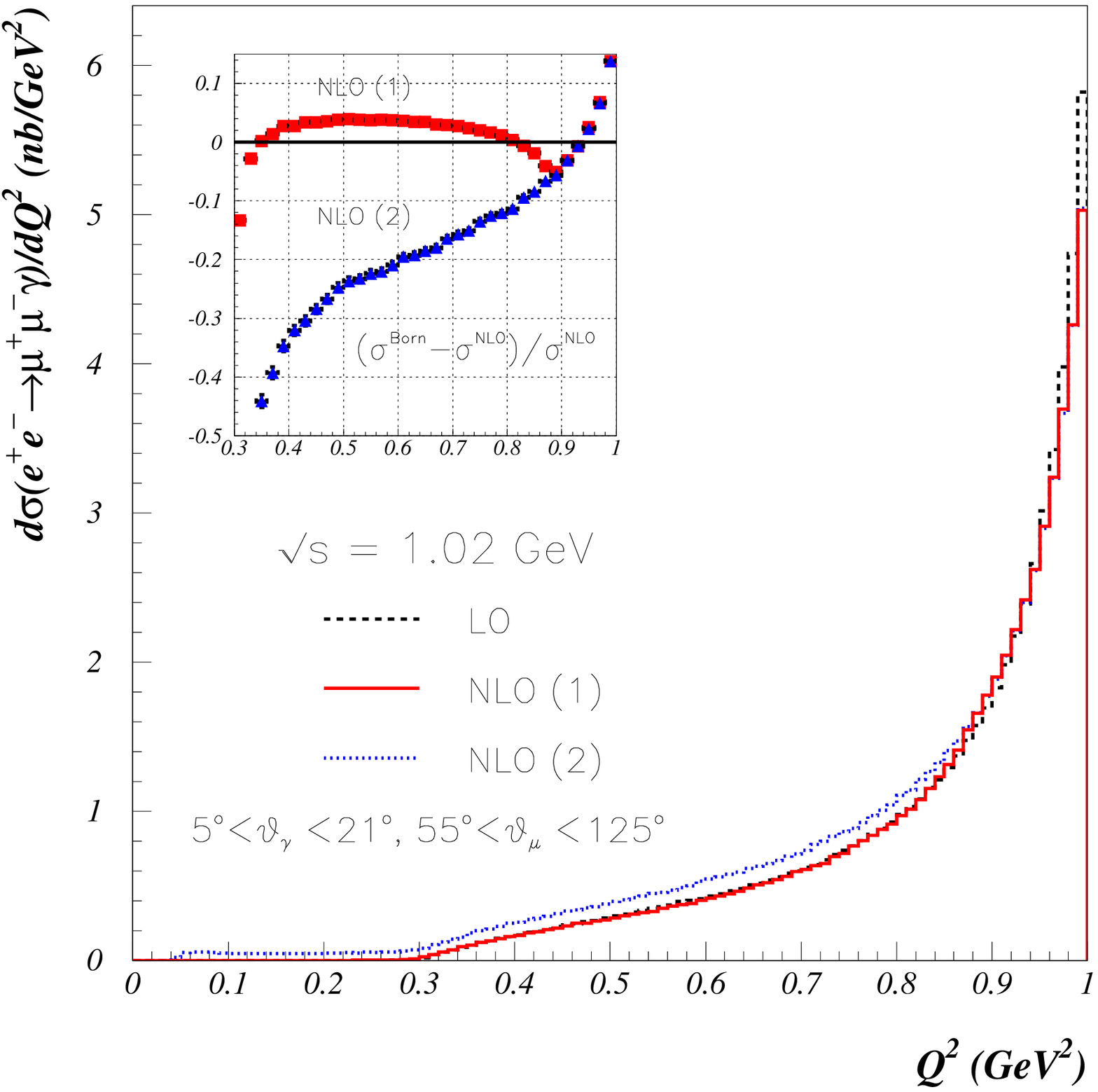,width=8.5cm} 
\end{center}
\caption{Differential cross section for the process 
$e^+ e^- \rightarrow \mu^+ \mu^- \gamma$ at NLO for $\sqrt{s}=1.02$~GeV.  
Only initial state radiation. The relative size of the correction to the 
LO result (dashed) is shown in the small inset. 
Same cuts as in figure~\ref{fig:nlo1gev}, with the pions replaced by muons.}
\label{fig:mnlo1gev}
\end{figure}

\begin{figure}
\begin{center}
\epsfig{file=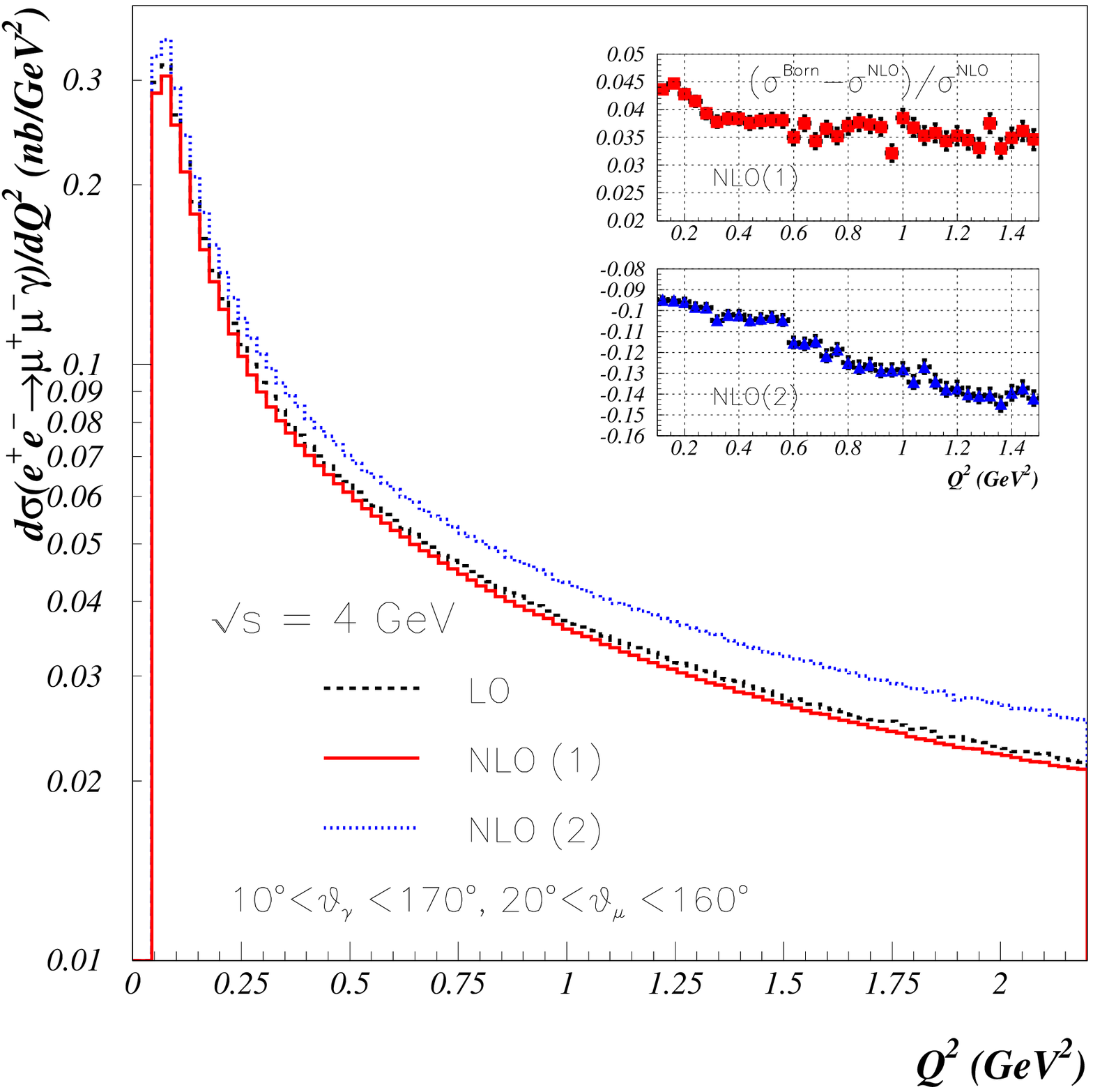,width=8.5cm} 
\end{center}
\caption{Differential cross section for the process 
$e^+ e^- \rightarrow \mu^+ \mu^- \gamma$ at NLO for $\sqrt{s}=4$~GeV.  
Only initial state radiation. The relative size of the correction to the 
LO result (dashed) is shown in the small inset. 
Same cuts as in Fig.~\ref{fig:nlo4gev}, with the pions replaced by muons.}
\label{fig:mnlo4gev}
\end{figure}

\begin{figure}
\begin{center}
\epsfig{file=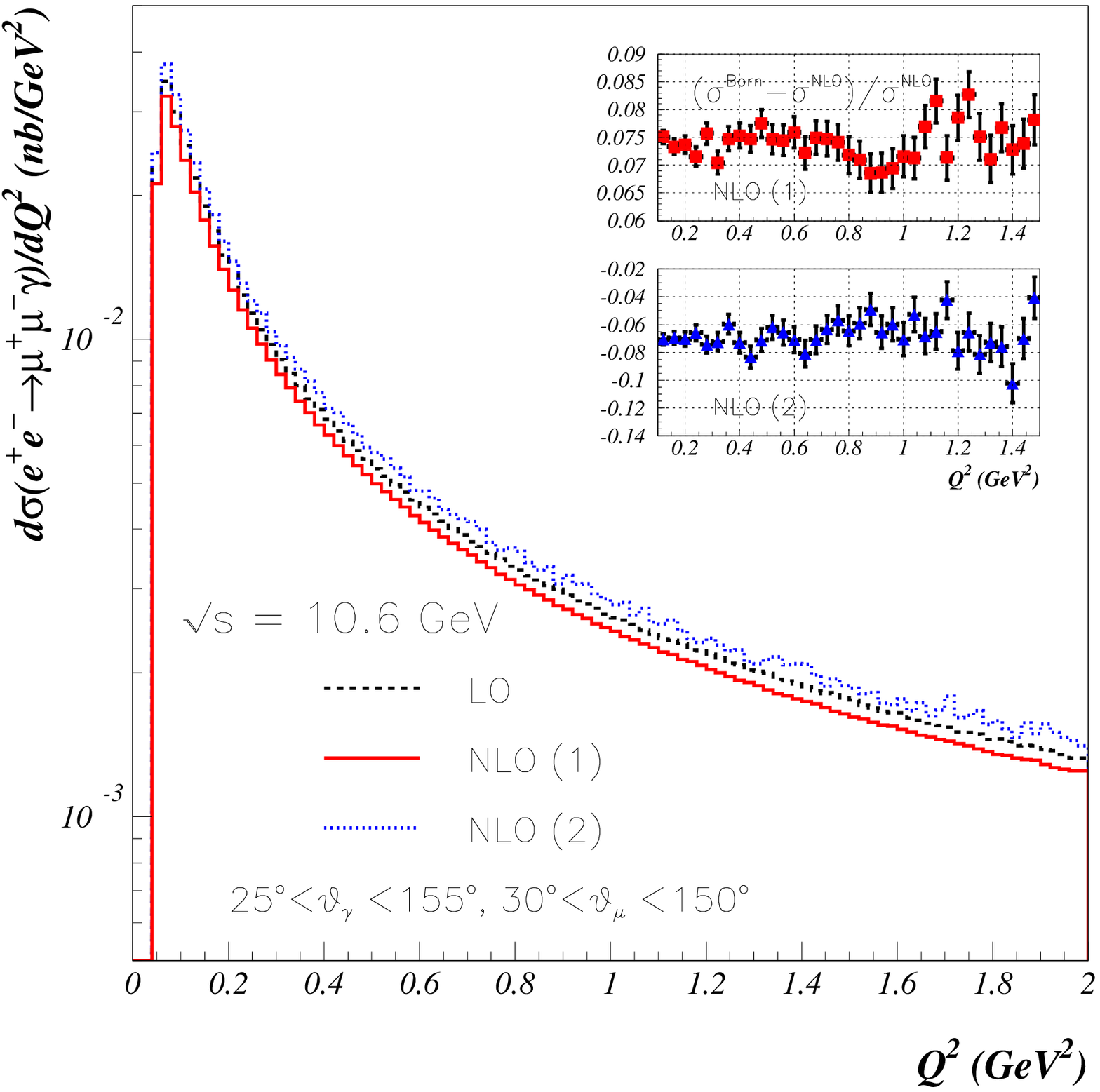,width=8.5cm} 
\end{center}
\caption{Differential cross section for the process 
$e^+ e^- \rightarrow \mu^+ \mu^- \gamma$ at NLO for $\sqrt{s}=10.6$~GeV.
Only initial state radiation. The relative size of the correction to the 
LO result (dashed) is shown in the small inset. 
Same cuts as in Fig.~\ref{fig:nlo10gev}, with the pions replaced by muons.}
\label{fig:mnlo10gev}
\end{figure}

\begin{figure}
\begin{center}
\epsfig{file=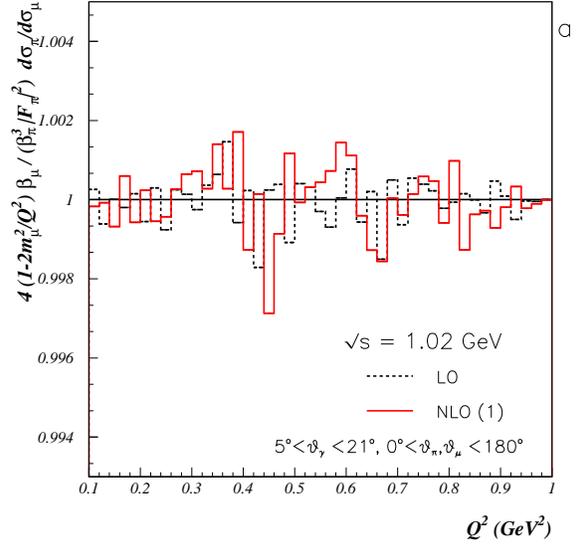,width=8cm}
\epsfig{file=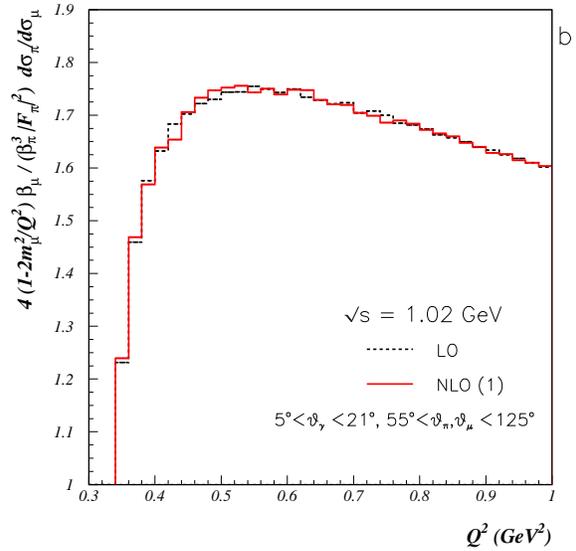,width=8cm}  
\end{center}
\caption{Ratio between pion and muon yields, after dividing through 
their respective R-ratio. Fig.~a: no cuts on pion and muon angles.
Fig.~b: with angular cuts on pion and muon angles.}
\label{fig:pipimumu}
\end{figure}

Inclusion of muon production in the program is straightforward. The 
results for the total cross section are listed in Table~\ref{tab:mxsec},
the differential distributions for the three cms energies in 
Figs.~\ref{fig:mnlo1gev} -- \ref{fig:mnlo10gev}. The radiative muon 
cross section could be used for a calibration of the pion yield.
A number of radiative corrections are expected to cancel in the 
ratio. For this reason we consider the ratio between 
the pion and the muon yields, after dividing the 
former by $|F_{\pi} (Q^2)|^2 (1-4m_\pi^2/Q^2)^{3/2}$, 
the latter by  $4 (1+2m_{\mu}^2/Q^2) \sqrt{1-4m_{\mu}^2/Q^2}$. 
In Fig.~\ref{fig:pipimumu}a we consider the full angular range for pions 
and muons, with $\theta_\gamma$ between $5^\circ$ and $21^\circ$.
Clearly all radiative corrections and kinematic effects disappear,
up to statistical fluctuations, in the leading order program as well 
as after inclusion of the NLO corrections.

In Fig.~\ref{fig:pipimumu}b an additional cut on pion and muon angles 
has been imposed. As demonstrated in Fig.~\ref{fig:pipimumu}b, the ratio 
differs from unity once (identical) angular cuts are imposed on pions 
and muons, a consequence of their different respective angular distribution.
To derive the pion form factor from the ratio between pion and muon 
yields, this effect has to be incorporated. However, the correction 
function shown in Fig.~\ref{fig:pipimumu}b is independent from the form factor, 
and hence universal. It can be obtained from the present program in a 
model-independent way (ignoring FSR for the moment).


\begin{table}
\caption{Total cross section (nb) for initial state radiation in 
the process $e^+ e^- \rightarrow \mu^+ \mu^- \gamma$ at LO,
NLO (1) and NLO (2) with the cuts from Table~\ref{tab:cuts}, 
the pions being replaced by muons.}
\label{tab:mxsec}
\begin{center}
\begin{tabular}{lccc}
  & $\sqrt{s}=$1.02~GeV & 4~GeV & 10.6~GeV \\ \hline  
Born     &  0.8243(5) &  0.4690(6) &  0.003088(6) \\
NLO (1)  &  0.7587(5) &  0.4449(6) &  0.002865(6) \\ 
NLO (2)  &  0.8338(7) &  0.4874(14) &  0.00321(6) \\ 
\hline
\end{tabular}
\end{center}
\end{table}

\section{Conclusions}

The radiative return with tagged photons offers a unique opportunity 
for a measurement of $\sigma(e^+e^-\rightarrow hadrons)$ over a wide 
range of energies. The reduction of the cross section by the 
additional factor $\alpha/\pi$ is easily compensated by the high 
luminosity of the new $e^+ e^-$ colliders, specifically the $\phi$-,
charm- and $B$-factories at Frascati, Cornell, Stanford and KEK.

In the present work we presented a Monte Carlo generator that 
simulates this reaction to next-to-leading accuracy. The current 
version includes initial state radiation only and is limited to 
$\pi^+ \pi^- \gamma (\gamma)$ and $\mu^+ \mu^- \gamma (\gamma)$
as final states. The uncertainty from unaccounted higher order
ISR is estimated at around $0.5\%$. The dominant FSR contribution can 
be deduced from the earlier program EVA. Additional hadronic modes 
can be easily implemented in the present program. The modes with 
three and four pions are in preparation.

\section*{Acknowledgements}

We would like to thank G.~Cataldi, A.~Denig, S.~Di~Falco, W.~Kluge, 
S.~M\"uller, G.~Venanzoni and B.~Valeriani for reminding us constantly 
of the importance of this work for the experimental analysis
and A.~Fern\'andez for very useful discussions.
Work supported in part by BMBF under grant number 05HT9VKB0 and 
E.U. EURODA$\Phi$NE network TMR project FMRX-CT98-0169. 
H.C. is grateful for the support and the kind hospitality
of the Institut f{\"u}r Theoretische Teilchenphysik
of the Karslruhe University.

\appendix

\section{Helicity amplitudes}

\label{app:helicity}

In this appendix the full set of helicity amplitudes for the 
diagrams of Fig.~\ref{fig:real} is given. Notation and 
calculation procedure were outlined in section~\ref{sec:two}, 
which follows Refs.~\cite{Jegerlehner:2000wu,Kolodziej:1991pk}.

As an example, consider the amplitude of the 
first Feynman diagram in Fig.~\ref{fig:real}: 

\begin{align}
M_1 &= T_1^{\mu} J_{\mu} =  \frac{e^3}{Q^2}  \bar{v}(p_1)   
\ta{\varepsilon}^*(k_1)[\ta{k}_1-\ta{p}_1+m_e] \non \\ & \times
\ta{\varepsilon}^*(k_2)[\ta{k}_1+\ta{k}_2-\ta{p}_1+m_e] 
\gamma^{\mu} u(p_2) J_{\mu} \non \\ & \times
\frac{1}{(2 k_1 \cdot p_1)(2 k_1 \cdot k_2 - 2 k_1 \cdot p_1
- 2 k_2 \cdot p_1)}~.
\end{align} 
Using the Dirac equation, after some 
algebra, the following expression is obtained
\begin{align}
M_1 &=  \frac{e^3}{Q^2} \bigg( 
v_I^{\dagger}(p_1) 
[\varepsilon^*(k_1)^- k_1^+ - 2\varepsilon^*(k_1)\cdot p_1] \non \\ & \times   
\varepsilon^*(k_2)^- [2p_2\cdot J - Q^+ J^-] u_I(p_2) \non \\ &
+ v_{II}^{\dagger}(p_1)
[\varepsilon^*(k_1)^+ k_1^- - 2\varepsilon^*(k_1)\cdot p_1] \non \\ & \times 
\varepsilon^*(k_2)^+ [2p_2\cdot J - Q^- J^+] u_{II}(p_2)
\bigg) \non \\ & \times
\frac{1}{(2 k_1 \cdot p_1)(2 k_1 \cdot k_2 - 2 k_1 \cdot p_1
- 2 k_2 \cdot p_1)}~.  
\end{align} 
Similar expressions are found for the amplitudes of the other diagrams
 and the complete matrix element can be written in a simple form (\ref{a1}),
 where the matrices \(A\left(\lambda_1,\lambda_2\right)\) and 
 \(B\left(\lambda_1,\lambda_2\right)\) are defined as:
\begin{eqnarray}
 A &=& - \frac{e^3}{4Q^2}\biggl(
 \frac{a_1  a_3}{(k_2 \cdot p_1)N_1}
 +\frac{a_5  a_3}{(k_1 \cdot p_1)N_1} 
  + \frac{a_5 a_7}{(k_1 \cdot p_1) (k_2 \cdot p_2)} \nonumber \\
 &&+ \frac{a_1 a_9}{(k_2 \cdot p_1) (k_1 \cdot p_2)}   
 + \frac{a_{11} a_7}{(k_2 \cdot p_2)N_2}
 + \frac{a_{11}a_9}{(k_1 \cdot p_2)N_2} \biggr) \ ,
 \nonumber \\
 B &=&  -\frac{e^3}{4Q^2}\biggl(
 \frac{a_2 a_4}{(k_2 \cdot p_1)N_1}
 +\frac{a_6 a_4}{(k_1 \cdot p_1)N_1}
  + \frac{a_6 a_8}{(k_1 \cdot p_1) (k_2 \cdot p_2)} \nonumber \\
 &&+\frac{ a_2 a_{10}}{(k_2 \cdot p_1) (k_1 \cdot p_2)}
 + \frac{a_{12} a_8}{(k_2 \cdot p_2)N_2}
 +\frac{a_{12}a_{10}}{(k_1 \cdot p_2)N_2}\biggr) \ \ ,\nonumber \\
\end{eqnarray}
where
\begin{eqnarray}
 N_1 &=&   k_1 \cdot p_1
 + k_2 \cdot p_1 - k_1 \cdot k_2~, \non \\
 N_2 &=&   k_1 \cdot p_2
 + k_2 \cdot p_2 - k_1 \cdot k_2~, \non \\
 a_1 &=& \varepsilon^*(k_2)^- k_2^+ - 2\varepsilon^*(k_2)\cdot p_1~, \non \\
 a_2 &=& \varepsilon^*(k_2)^+ k_2^- - 2\varepsilon^*(k_2)\cdot p_1~, \non \\
 a_3 &=& 2p_2\cdot J - Q^+J^-~, \non \\
 a_4 &=& 2p_2\cdot J - Q^-J^+~, \non \\
 a_5 &=& \varepsilon^*(k_1)^- k_1^+ - 2\varepsilon^*(k_1)\cdot p_1~, \non \\
 a_6 &=& \varepsilon^*(k_1)^+ k_1^- - 2\varepsilon^*(k_1)\cdot p_1~, \non \\
 a_7 &=&  2\varepsilon^*(k_2)\cdot p_2 - k_2^+\varepsilon^*(k_2)^-~, \non \\
 a_8 &=&  2\varepsilon^*(k_2)\cdot p_2 - k_2^-\varepsilon^*(k_2)^+~, \non \\
 a_9 &=&  2\varepsilon^*(k_1)\cdot p_2 - k_1^+\varepsilon^*(k_1)^-~, \non \\
 a_{10}
  &=&  2\varepsilon^*(k_1)\cdot p_2 - k_1^-\varepsilon^*(k_1)^+~, \non \\
 a_{11} &=&  J^-Q^+ - 2p_1\cdot J~, \non \\
 a_{12} &=&  J^+Q^- - 2p_1\cdot J~. \non \\
\end{eqnarray}

 The current \(J^\mu\) is defined by the \eq{hc} for the \(\pi^+\pi^-\)
 final state, while for the \(\mu^+\mu^-\) one it is defined as follows:
\begin{eqnarray}
 J^\mu(\lambda_1,\lambda_2) = i e \bar u(q_1,\lambda_1)\gamma^\mu 
 v(q_2,\lambda_2) \ \ ,
\end{eqnarray}

\noindent
where \(q_1,\lambda_1\) (\(q_2,\lambda_2\)) are four-momentum and helicity
of \(\mu^-\) (\(\mu^+\)).

\section{Phase space}

\label{app:phasespace}

The generation of the multiparticle phase space is based on the 
following Lorentz-invariant representation: 
\begin{align} 
& d \Phi_{m+n}(p_1,p_2;k_1,\cdot,k_m,q_1,\cdot,q_n)
= \non \\ & \qquad d \Phi_m(p_1,p_2;Q,k_1,\cdot,k_m) d \Phi_n(Q;q_1,\cdot,q_n) 
\frac{dQ^2}{2\pi}~,
\end{align}
where $p_1$ and $p_2$ are the four-momenta of the initial particles,
$k_1 \ldots k_m$ are the four momenta of the emitted photons and 
$q_1 \ldots q_n$, with $Q = \sum q_i$, label the four-momenta of the 
hadrons. 

When two pions are produced in the final state, the hadronic part of 
phase space is given by
\begin{equation}
d \Phi_2(Q;q_1,q_2) = \frac{\sqrt{1-\frac{4 m_{\pi}^2}{Q^2}}}{32 \pi^2} 
d \Omega~,
\end{equation}
where $d \Omega$ is the solid angle of one of the pions at, for instance, 
the $Q^2$ rest frame.
 
One single photon emission is described by the corresponding 
leptonic part of phase space 
\begin{equation}
d \Phi_2(p_1,p_2;Q,k_1) = \frac{1-q^2}{32 \pi^2} d \Omega_1~,
\end{equation}
with $q^2=Q^2/s$ and $d \Omega_1$ is the solid angle of the emitted 
photon at the $e^+ e^-$ rest frame. The polar angle $\theta_1$
is defined with respect to the positron momentum $p_1$.
In order to make the Monte Carlo generation more efficient,
the following substitution is performed:
\begin{align}
\cos \theta_1 = \frac{1}{\be} \tanh(\be \; t_1) ~, \quad
t_1 = \frac{1}{2\be} \log \frac{1+\be \cos \theta_1}{1-\be \cos \theta_1}~,
\label{t1}
\end{align}
with $\be = \sqrt{1-4m_e^2/s}$,
which accounts for the collinear emission peaks  
\begin{equation}
\frac{d \cos \theta_1}{1-\be^2 \cos^2 \theta_1} = dt_1~.
\end{equation}
Then, the azimuthal angle and the new variable $t_1$ are generated flat. 

Consider now the emission of two real photons. 
In the cms of the initial particles, the four-momenta 
of the positron, the electron and the two emitted photons are given by 
\begin{align}
p_1 &= \frac{\sqrt{s}}{2}(1,0,0,\be)~, \qquad 
p_2 = \frac{\sqrt{s}}{2}(1,0,0,-\be)~, \non \\
k_1 &= w_1 \sqrt{s} (1,\sin \theta_1 \cos \phi_1,\sin \theta_1 \sin \phi_1,
\cos \theta_1)~, \non \\
k_2 &= w_2 \sqrt{s} (1,\sin \theta_2 \cos \phi_2,\sin \theta_2 \sin \phi_2,
\cos \theta_2)~,
\end{align}
respectively. The polar angles $\theta_1$ and $\theta_2$ are
defined again with respect to the positron momentum $p_1$.
Both photons are generated with energies larger than 
the soft photon cutoff: $w_i>w$ with $i=1,2$. At least 
one of these should exceed the minimal detection energy: 
$w_1 > E_{\gamma}^{min}/s$ or $w_2 > E_{\gamma}^{min}/s$.
In terms of the solid angles $d \Omega_1$ and $d \Omega_2$ of the 
two photons and the normalized energy of one of them, e.g. $w_1$,
the leptonic part of phase space reads 
\begin{align}
d \Phi_3(p_1,p_2;Q,k_1,k_2) &= \frac{1}{2!} \;\frac{s}{4(2\pi)^5} 
\non \\ \times &  \frac{w_1 w_2^2}{1-q^2-2w_1}
\; dw_1 \; d\Omega_1 \; d\Omega_2~,
\end{align}
where the limits of the phase space are defined by the constraint 
\begin{equation}
q^2 = 1-2(w_1+w_2)+2w_1 w_2(1-\cos \chi_{12})~,
\label{dpslimits}
\end{equation}
with $\chi_{12}$ being the angle between the two photons
\begin{equation}
\cos \chi_{12} = \sin \theta_1 \sin \theta_2 \cos (\phi_1 - \phi_2) 
+ \cos \theta_1 \cos \theta_2~.
\end{equation}

Again, the matrix element squared contains several peaks, soft and 
collinear, which should be softened by choosing suitable substitutions
in order to achieve an efficient Monte Carlo generator.
The leading behaviour of the matrix element squared is given 
by $1/(y_{11} \; y_{12} \; y_{21} \; y_{22})$, where  
\begin{equation}
y_{ij} = \frac{2 k_i \cdot p_j}{s} = w_i (1 \mp \be \cos \theta_i)~.
\end{equation}
In combination with the leptonic part of phase space, we have
\begin{align}
& \frac{d\Phi_3(p_1,p_2;Q,k_1,k_2)}{y_{11} \; y_{12} \; y_{21} \; y_{22}}
\sim
\frac{dw_1}{w_1(1-q^2-2w_1)} \; \non \\ & \qquad \times 
\frac{d\Omega_1}{1-\be^2 \cos^2{\theta_1}}
\; \frac{d\Omega_2}{1-\be^2 \cos^2{\theta_2}}~.
\end{align}
The collinear peaks are then flattened with the help of \eq{t1}, with 
one change of variables for each photon polar angle. The remaining soft 
peak, $w_1 \rightarrow w$, is reabsorbed with the following substitution
\begin{align}
w_1 = \frac{1-q^2}{2+e^{-u_1}}~, \quad u_1 = \log \frac{w_1}{1-q^2-2w_1}~,
\end{align}
or
\begin{align}
\frac{d w_1}{w_1(1-q^2-2w_1)} = \frac{d u_1}{1-q^2}~,
\end{align}
where the new variable $u_1$ is generated flat.



\begin{thebibliography}{99}

\bibitem{Brown:2001mg}
H.~N.~Brown {\it et al.}  [Muon $g-2$ Collaboration],
Phys.\ Rev.\ Lett.\  {\bf 86} (2001) 2227
[hep-ex/0102017].

\bibitem{Hughes:1999fp}
V.~W.~Hughes and T.~Kinoshita,
Rev.\ Mod.\ Phys.\ {\bf 71} (1999) S133.

\bibitem{hadronicmuon}
S.~Eidelman and F.~Jegerlehner,
Z.\ Phys.\ C {\bf 67} (1995) 585
[hep-ph/9502298].
D.~H.~Brown and W.~A.~Worstell,
Phys.\ Rev.\ D {\bf 54} (1996) 3237
[hep-ph/9607319].
M.~Davier and A.~H\"ocker,
Phys.\ Lett.\ B {\bf 435} (1998) 427
[hep-ph/9805470].
S.~Narison,
Phys.\ Lett.\ B {\bf 513} (2001) 53
[hep-ph/0103199].
F.~Jegerlehner,
[hep-ph/0104304].
J.~F.~De Troc\'oniz and F.~J.~Yndur\'ain,
[hep-ph/0106025].

\bibitem{runningQED}
H.~Burkhardt and B.~Pietrzyk,
Phys.\ Lett.\ B {\bf 513} (2001) 46.
A.~D.~Martin, J.~Outhwaite and M.~G.~Ryskin,
[hep-ph/0012231];
Phys.\ Lett.\ B {\bf 492} (2000) 69
[hep-ph/0008078].
J.~Erler,
Phys.\ Rev.\ D {\bf 59} (1999) 054008
[hep-ph/9803453].
J.~H.~K\"uhn and M.~Steinhauser,
[hep-ph/0109084];
Phys.\ Lett.\ B {\bf 437} (1998) 425
[hep-ph/9802241].

\bibitem{Binner:1999bt}
S.~Binner, J.~H.~K\"uhn and K.~Melnikov,
Phys.\ Lett.\  {\bf B459} (1999) 279
[hep-ph/9902399].

\bibitem{Melnikov:2000gs}
K.~Melnikov, F.~Nguyen, B.~Valeriani and G.~Venanzoni,
Phys.\ Lett.\  {\bf B477} (2000) 114 
[hep-ph/0001064].

\bibitem{Czyz:2000wh}
H.~Czy\.z and J.~H.~K\"uhn, 
Eur.\ Phys.\ J.\ C {\bf 18} (2001) 497
[hep-ph/0008262].

\bibitem{Spagnolo:1999mt}
S.~Spagnolo,
Eur.\ Phys.\ J.\ C {\bf 6} (1999) 637.

\bibitem{Khoze:2001fs}
V.~A.~Khoze, M.~I.~Konchatnij, N.~P.~Merenkov, G.~Pancheri, L.~Trentadue and O.~
N.~Shekhovzova,
Eur.\ Phys.\ J.\ C {\bf 18} (2001) 481
[hep-ph/0003313].

\bibitem{Akhmetshin:1999uj}
R.~R.~Akhmetshin {\it et al.}  [CMD-2 Collaboration],
[hep-ex/9904027].

\bibitem{Hoefer:2001mx}
A.~H\"ofer, J.~Gluza and F.~Jegerlehner,
[hep-ph/0107154].

\bibitem{Aloisio:2001xq}
A.~Aloisio {\it et al.}  [KLOE Collaboration],
[hep-ex/0107023].

\bibitem{Denig:2001ra}
A.~Denig {\it et al.}  [KLOE Collaboration],
eConf {\bf C010430} (2001) T07
[hep-ex/0106100].

\bibitem{Adinolfi:2000fv}
M.~Adinolfi {\it et al.}  [KLOE Collaboration],
[hep-ex/0006036].

\bibitem{babar}
E.~P.~Solodov  [BABAR collaboration],
eConf {\bf C010430} (2001) T03
[hep-ex/0107027].

\bibitem{Rodrigo:2001jr}
G.~Rodrigo, A.~Gehrmann-De Ridder, M.~Guilleaume and J.~H.~K\"uhn,
Eur.\ Phys.\ J.\ C {\bf 22} (2001) 81
[hep-ph/0106132].

\bibitem{Rodrigo:2001cc}
G.~Rodrigo,
Acta Phys.\ Polon.\ B {\bf 32} (2001) 3833 [hep-ph/0111151].


\bibitem{Kuhn:1990ad}
J.~H.~K\"uhn and A.~Santamaria,
Z.\ Phys.\ C {\bf 48} (1990) 445.

\bibitem{BHS:1986}
M.~B\"ohm, H.~Spiesberger and W.~Hollik,
Fortsch.\ Phys.\  {\bf 34} (1986) 687.

\bibitem{Jegerlehner:2000wu}
F.~Jegerlehner and K.~Ko\l odziej,
Eur.\ Phys.\ J.\ C {\bf 12} (2000) 77
[hep-ph/9907229].

\bibitem{Kolodziej:1991pk}
K.~Ko\l odziej and M.~Zra\l ek,
Phys.\ Rev.\ D {\bf 43} (1991) 3619.

\bibitem{structure}
E.~A.~Kuraev and V.~S.~Fadin,
Sov.\ J.\ Nucl.\ Phys.\ {\bf 41} (1985) 466.
G.~Altarelli and G.~Martinelli,
in {\it Physics at LEP}, CERN Report 86-06, 
eds. J.~Ellis and R.~Peccei (CERN, Geneva, February 1986).
O.~Nicrosini and L.~Trentadue,
Phys.\ Lett.\ {\bf B196} (1987) 551.

\bibitem{Caffo:1997yy}
M.~Caffo, H.~Czy\.z and E.~Remiddi,
Nuovo Cim.\ A {\bf 110} (1997) 515
[hep-ph/9704443];
Phys.\ Lett.\  {\bf B327} (1994) 369.

\bibitem{Kniehl:1988id}
B.~A.~Kniehl, M.~Krawczyk, J.~H.~K\"uhn and R.~G.~Stuart,
Phys.\ Lett.\ B {\bf 209} (1988) 337.








\end{thebibliography}
\end{document}